%% file: main.tex
\documentclass[sigconf]{acmart}
\pdfoutput=1

\renewcommand\footnotetextcopyrightpermission[1]{}

\usepackage{amsmath}
\usepackage{mathtools}
\usepackage{mathpartir}
\usepackage[skip=2.5pt]{caption}
\usepackage{subcaption}
\usepackage{listings}
\usepackage{multirow}
\usepackage[ruled,linesnumbered]{algorithm2e}
\usepackage{enumitem}
\usepackage{threeparttable}
\usepackage[compact]{titlesec}
\usepackage{pifont}
\usepackage{bbding}
\usepackage{pdfpages}

\newcommand{\bv}[1]{\ensuremath{\text{Vec}\langle #1 \rangle}}
\newcommand{\rad}{\text{URA}}
\newcommand{\ura}{\text{URA}}
\newcommand{\wra}{\text{WRA}}
\newcommand{\sdd}{\text{SDD}}
\newcommand{\cst}{\text{CST}}
\newcommand{\sid}{\text{SID}}
\newcommand{\ukd}{\text{WRA}}
\newcommand{\tool}{\textsc{CaType}}
\newcommand{\subtype}{\ensuremath{\leq:}}

\newcommand{\parh}[1]{\noindent\textbf{#1}}

\newcommand{\F}{Fig.}

\newcommand{\T}{Table}
\renewcommand{\S}{Sec.}

\newcommand{\rev}[1]{#1}

\newcommand\DejaVuttfamily{%
  \fontfamily{DejaVuSansMono-TLF}\selectfont
}

\lstdefinestyle{base}{
  moredelim=**[is][\color{red}]{@}{@},
  escapeinside={<@}{@>}
}

\lstdefinelanguage
   [x64]{Assembler}     
   [x86masm]{Assembler} 
   {morekeywords={CDQE,CQO,CMPSQ,CMPXCHG16B,JRCXZ,LODSQ,MOVSXD, %
                  POPFQ,PUSHFQ,SCASQ,STOSQ,IRETQ,RDTSCP,SWAPGS, %
                  rax,rdx,rcx,rbx,rsi,rdi,rsp,rbp, %
                  r8,r8d,r8w,r8b,r9,r9d,r9w,r9b}} 

\begin{document}

\title{Cache Refinement Type for Side-Channel Detection of Cryptographic Software}

\author{Ke Jiang}
\affiliation{
  \institution{Nanyang Technological University}
  \city{Singapore}
  \country{Singapore}}
\email{ke006@e.ntu.edu.sg}

\author{Yuyan Bao}
\affiliation{
  \institution{University of Waterloo}
  \city{Waterloo}
  \state{Ontario}
  \country{Canada}}
\email{yuyan.bao@uwaterloo.ca}

\author{Shuai Wang}
\authornote{Corresponding authors}
\affiliation{
  \institution{Hong Kong University of Science and Technology}
  \city{Hong Kong}
  \country{China}}
\email{shuaiw@cse.ust.hk}

\author{Zhibo Liu}
\affiliation{
  \institution{Hong Kong University of Science and Technology}
  \city{Hong Kong}
  \country{China}}
\email{zliudc@cse.ust.hk}

\author{Tianwei Zhang}
\affiliation{
  \institution{Nanyang Technological University}
  \city{Singapore}
  \country{Singapore}}
\authornotemark[1]
\email{tianwei.zhang@ntu.edu.sg}

\input{0_abstract}




\maketitle

\input{1_introduction}
\input{2_preliminaries}
\input{3_overview}

\input{4_design}
\input{5_implementation}
\input{6_evaluation}
\input{7_discussion}
\input{8_relatedwork}
\input{9_conclusion}

\begin{acks}
We would like to thank the anonymous reviewers for their constructive feedback. This work has been supported in part by Singapore National Research Foundation under its National Cybersecurity R\&D Programme (NCR Award NRF2018 NCR-NCR009-0001), Singapore Ministry of Education (MOE) AcRF Tier 1 RS02/19, NTU Start-up grant.
\end{acks}

\bibliographystyle{ACM-Reference-Format}
\bibliography{reference}

\appendix
\input{10_appendices}


\end{document}

%% file: 0_abstract.tex
\begin{abstract}

Cache side-channel attacks exhibit severe threats to software security and
privacy, especially for cryptosystems. In this paper, we propose \tool, a novel
refinement type-based tool for detecting cache side channels in crypto software.
Compared to previous works, \tool\ provides the following advantages: (1) For the 
first time \tool\ analyzes cache side channels using refinement type
over x86 assembly code. It reveals several significant and effective
enhancements with refined types, including bit-level granularity tracking,
distinguishing different effects of variables, precise type inferences, and high
scalability. (2) \tool\ is the first static analyzer for crypto libraries in 
consideration of blinding-based defenses.
(3) From the perspective of implementation, \tool\ uses cache layouts of
potential vulnerable control-flow branches rather than cache states to suppress
false positives.
We evaluate \tool\ in identifying side channel vulnerabilities in real-world
crypto software, including RSA, ElGamal, and (EC)DSA from OpenSSL and Libgcrypt.
\tool\ captures all known defects, detects previously-unknown vulnerabilities,
and reveals several false positives of previous tools. In terms of performance,
\tool\ is 16$\times$ faster than CacheD and 131$\times$ faster than CacheS when 
analyzing the same libraries. These evaluation results confirm the capability of 
\tool\ in identifying side channel defects with great precision, efficiency, and 
scalability.

\end{abstract}

%% file: 1_introduction.tex
\section{Introduction}
\label{sec:introduction}

Cache-based side channels have demonstrated serious threats to crypto
algorithms, such as the symmetric cipher AES~\cite{percival2005cache,
osvik2006cache}, the asymmetric cipher RSA~\cite{zhang2012cross, yarom2014flush,
liu2015last}, and the digital signature (EC)DSA~\cite{yarom2014recovering,
ryan2019return, aranha2020ladderleak}. The essence of these cache
attacks is the interference of program memory accesses toward cache units,
where secret-dependent memory accesses or program branches leave distinguishable
footprints in cache units. Thus, identifying and removing cache interference can eliminate side channel leakage.

Designing novel security-aware cache architectures may eliminate adversarial
interference. Prior research relies mostly on two strategies, namely
partitioning-based and randomization-based approaches. Strong isolation is
achieved in partition isolated caches~\cite{wang2007new, domnitser2012non} by
physically partitioning the shared cache into multiple zones for applications of
various security levels. In contrast,~\cite{wang2007new, wang2008novel,
qureshi2018ceaser, werner2019scattercache} obscure adversary observations by
randomizing the cache states. Although it is envisaged that these architectures
will eliminate interference and secure programs that run on top of them, recent
works show that these randomization-based caches may be still vulnerable to
cache side channels~\cite{song2021randomized, purnal2021systematic}. Also, these
new cache designs achieve security promise at the expense of performance.
Besides, they are not yet ready for commercial use due to extra cost in chip
circuit manufacturing.

Software-based mitigation of cache side channels appears increasingly viable.
However, manually detecting vulnerable crypto code takes specialized knowledge,
which drastically restricts normal developers from analyzing and patching their
crypto software. With the fast development of more efficient crypto software
under various usage scenarios, launching timely side channel analysis becomes
even more challenging. With this regard, developing a general, automated, and
efficient analytic tool for detecting cache side channels is
receiving broad attention from both academics and industry. Recent
works~\cite{doychev2013cacheaudit, doychev2017rigorous, wang2017cached,
weiser2018data, brotzman2019casym, wang2019identifying} serve as examples of
this. In general, these works construct constraints through symbolic modeling of
program states and cache accesses. Then, constraint solving techniques (e.g.,
Z3~\cite{z3}) are employed to check the satisfiability of constraints and decide
whether the program is vulnerable to cache side channels. While these automated
methods have made concrete progress in discovering cache side channels in
real-world cryptosystems, they still face a number of obstacles.

\noindent \textbf{Challenge 1:}~\textit{Software-based analysis needs to address
precision issues and be scalable to production crypto libraries.}
CacheAudit~\cite{doychev2013cacheaudit} and its
extension~\cite{doychev2017rigorous} calculate the upper bound of information
leakage by counting all possible final cache states via abstract
interpretation~\cite{cousot77abstract}. However, estimating worst-case leakage
bound may not reflect the reality. Moreover, CacheAudit cannot pinpoint
what/where the vulnerability is, prohibiting the debugging/fixing of analyzed
code. Using symbolic execution, CaSym~\cite{brotzman2019casym} distinguishes two
different cache states resulting from secret variants. Though CaSym covers
multiple paths, it suffers from path explosion and is less scalable.
CacheS~\cite{wang2019identifying}, likely the most scalable static tool in this
field, also uses abstract interpretation. It achieves higher scalability due to
modeling secret/non-secret semantics with symbolic formulas of different
granularity. Dynamic approaches, in contrast, analyze concrete execution traces
to track program states and pinpoint side channels. CacheD~\cite{wang2017cached}
detects secret-dependent memory accesses via symbolic execution, while not
considering secret-dependent branches. DATA~\cite{weiser2018data} considers both
memory access leaks and branch leaks through differentiating address traces.
Existing dynamic methods, though manifest relatively improved
scalability, may still be slow to analyze production crypto libraries (due to
the usage of constraint solving) or require many well-chosen inputs to induce
distinct observations.

\noindent \textbf{Challenge 2:}~\textit{Cache models adopted by software
analyzers have an effect on the scalability and detection granularity.} Relying
on concrete cache replacement policies (e.g., LRU, FIFO, and RLRU), CacheAudit
precisely describes a program been executed on the expected architecture, at the
cost of scalability due to architectural complexity. CaSym uses high-level
abstract cache models (i.e., infinite and age models) to achieve higher analysis
scalability. It uses the array index to compute the accessed cache locations.
However, these abstract models have granularity issues: there is a gap between
the array index and the cache location in realistic architectures. At the other
extreme, a much simplified cache model is shared by~\cite{doychev2017rigorous,
wang2017cached, weiser2018data, wang2019identifying, bao2021abacus},
where an architectural-independent model is used to detect cache side
channels. Though this model is realistic and efficient, performing analysis at
such granularity results in false positives, as will be discussed in this
paper.

\noindent \textbf{Challenge 3:}~\textit{Supporting a comprehensive analysis of
crypto software rather than some specific defects in sensitive code fragments.}
For instance, CacheD omits the analysis of secret-dependent program
branches. Moreover, modern crypto libraries extensively use randomization
schemes like binding to mitigate side channels, whose effectiveness (and
remaining leaks) have not been analyzed by previous tools. \rev{Supporting
randomization is inherently hard for previous static (abstract
interpretation-based) tools~\cite{doychev2013cacheaudit, doychev2017rigorous,
wang2019identifying}, requiring new abstract domains, new abstract operators,
and soundness proofs. Meanwhile, modeling randomization is also costly for
approaches that use constraint solvers, as it demands to iterate blinding
quantifiers~\cite{wang2017cached, wang2019identifying, brotzman2019casym}.}
\cite{weiser2018data} conceptually differentiates traces derived from
blinding-involved computations, but it overlooks the complex computations
involving blinding in production cryptosystems, which may contain new attack
vectors.

The aforementioned obstacles incentive the design of \tool, an automated,
precise, and efficient cache side-channel analysis tool. \tool\ is scalable and
capable of analyzing large-scale, complex crypto software. \tool\
follows~\cite{wang2017cached, weiser2018data} to log execution traces of crypto
software \rev{and performs trace-based type inference on the logged traces.
It features a novel refinement type system that enables tracking
program variables in the bit-level representation. Different from previous
constraint solving-based approaches that are inherently costly, our sound type
system guarantees fine-grained secret tracking and side channel detection with largely improved efficiency.}
Lastly, \tool\ comprehensively models randomization-based
mitigation schemes adopted in modern crypto software. \rev{It allocates
specific refined types for differentiating the responsibilities of (secret or
randomized) variables, enabling precise information flow tracking under the
presence of randomization.} In sum, we make the following contributions:

\begin{itemize}[leftmargin=10pt]

\item[$\bullet$] \rev{Conceptually, for the first time, cache side channels are 
analyzed using refinement type techniques. We establish our novel refinement
type system directly over x86 assembly code, and formulate cache side channels
over refined types.}

\item[$\bullet$] \rev{Technically, \tool\ features several important and
effective enhancements compared with prior tools on the basis of refinement type
system, including bit-level granularity tracking, distinguishing different
effects of variables, precise type inferences, and much higher scalability.
\tool\ takes into account randomization-based defenses using specific refined
types, and uses novel cache layouts to suppress potential false
positives.}


\item[$\bullet$] \rev{Empirically, we evaluate \tool\ to uncover side channel
vulnerabilities among real-world crypto libraries. \tool\ captures all known design flaws,
identifies unknown flaws, and reveals several false positives in existing tools.
\tool\ is 16$\times$ faster than CacheD and 131$\times$ faster than CacheS,
demonstrating its high applicability toward production crypto software.}

\end{itemize}



%% file: 2_preliminaries.tex
\section{Preliminaries}
\label{sec:background}

\subsection{Refinement Type Systems}
\label{subsec:refinement}

\rev{A type system is a well-established formal system comprising a set of rules
that assigns types to terms in a programming
language~\cite{cardelli1996type, pierce2002types}. 
For example, C language contains a basic type system,
where types (e.g., \texttt{int}, \texttt{double}, and \texttt{int*}) give
\textit{meaning} to data in the memory or registers. Modern C compilers can
feature basic type checking rules to detect invalid operations, e.g., when a
variable of double is used as \texttt{int*} (for pointer dereference), an error
is thrown at the compilation time.}

\rev{Type systems are widely-used in language-based security
research~\cite{zhang2012language} like tracking secure information flow.
In those systems, the types of variables and expressions are attached with annotations that specify confidentiality policies enforcing the use of the typed data.
For instance, two type annotations $H$ and $L$ are used to denote high and low
security sensitivity of data. To detect the violation of confidentiality policy,
a set of type rules is defined to check if the two classified sets of data
interfere with each other.}

\rev{Refinement types~\cite{jhala2021refinement} 
extend standard type annotations with predicates that confine the use of the values described by the type. Typically, a variable $x$'s refinement type can
be defined in the form of $x:T\{v:P\}$, where $T$ is a basic type and $P$ is the
associated predicate. For example, a non-negative integer variable $x$ is
represented as $x:int\{v:0 \le v\}$, where predicate $0 \le v$ refines the basic
type $int$ by specifying that the integer must be greater than or equal to zero.
With well-defined predicates, the refinement types can provide stronger guarantees. For example, the zero-division errors can be alerted
at the compilation time when the predicate $N \ge 0$ indicates that the divisor may be
zero. Meanwhile, one can elaborately specify security policies over the
refinement types to verify software security
vulnerabilities.~\cite{bhargavan2010modular, bengtson2011refinement,
bhargavan2013implementing, barthe2014probabilistic} are successful examples of adopting refinement type systems in high-level languages (e.g., F$^\ast$) to provide security guarantees in crypto infrastructures.
To our best knowledge, 
\tool\ is the first to employ refinement types over assembly code and for
cache side channel detection.}

\subsection{Cache Hierarchy}
\label{subsec:prel-cache}

Caches are incorporated into CPUs to accelerate process execution due to the
locality principle. In modern CPUs, each core (i.e., a processing unit on a CPU
chip) monopolizes an L1 cache and a L2 cache. All cores share a megabyte-size
LLC (Last-Level Cache). 
The access time for a cache hit is around tens of cycles. In contrast, the
latency will become much higher (usually hundreds of cycles) when a cache miss
occurs and the main memory has to be accessed.
Modern CPUs use a $W$-way set-associative cache. Different memory blocks may
reside on the same cache set, and each cache set is further divided into $W$
cache lines. 
Given an $N$-bit memory address, $S$-set cache with $L$ byte-size cache line,
the lowest $log_{2}L$ bits of the address represent the offset since continuous
memory blocks are cached together within one \texttt{load} instruction. The
middle $log_{2}S$ bits starting from bit $log_{2}L$ are used to locate the cache
set index. The upper part represents cache hit/miss tag bits.

\subsection{Cache Side Channels}
\label{subsec:cache-sc}

\rev{Cache poses threats of secret leakage, as program cache accesses may be
leveraged by adversaries to reconstruct confidential information. In this
section, we introduce two representative vulnerable code patterns,
\textit{secret-dependent branch condition (SDBC)} and \textit{secret-dependent
memory access (SDMA)}, via classic examples in RSA.}

\parh{Secret-Dependent Branch Condition (SDBC).}~\rev{\F~\ref{fig:channel1} shows
a simplified view of the square-and-multiply implementation of modular
exponentiation in RSA. $e_{i}$ (line 4) denotes a private key and decides if
line 5 is executed. By monitoring the L1 instruction cache (I-cache), attackers
are aware of the execution of line 5, and further reconstruct $e_i$ using
well-established cache attacks~\cite{liu2015last,yarom2014flush}.}

\parh{Secret-Dependent Memory Access (SDMA).}~\rev{Besides SDBC, SDMA also leads to
exploitations. Consider \F~\ref{fig:channel2}, where the sliding window modular
exponentiation algorithm initializes a precomputed array $g[i]$ (lines 1--3) to
accelerate the computation. When performing decryption, a window size key
$w_{i}$ (line 8) is used as the index to query the precomputed table $g[i]$. For
each for-loop (line 8), monitoring the accessed data cache (D-cache) line can
reveal certain bits in $w_{i}$ and gradually reconstruct the private
key~\cite{liu2015last}.}

\begin{figure}[!htbp]
\begin{minipage}[c]{0.4\linewidth}
    \begin{subfigure}[b]{0.9\linewidth}
    \centering
    \footnotesize
    \begin{tabular}{ll}
        &$1:\ x \leftarrow 1$\\
        &$2:\ \textbf{for}\ i \leftarrow \vert e \vert - 1\ downto\ 0$\\
        &$3:\ \quad x \leftarrow x^{2}\ mod\ m$\\
        &$4:\ \quad \textbf{if}\ e_{i} = 1\ \textbf{then}$\\
        &$5:\ \quad \quad \textcolor{blue}{x \leftarrow x \cdot b\ mod\ m}$\\
        &$6:\ \textbf{return}\ x$
    \end{tabular}
    \caption{Square-and-Multiply Exp.}
    \label{fig:channel1}
    \end{subfigure}
\end{minipage}\hspace{5pt}
\begin{minipage}[c]{0.55\linewidth}
    \begin{subfigure}[b]{0.9\linewidth}
    \centering
    \footnotesize
    \begin{tabular}{ll}
        &$1:\ g[0] \leftarrow b\ mod\ m$\\
        &$2:\ \textbf{for}\ j \leftarrow 1\ to\ 2^{S-1} - 1$\\
        &$3:\ \quad \ g[j] \leftarrow b^{2j + 1} \ mod\ m$\\
        &$4:\ x \leftarrow \textcolor{blue}{g[(w_{n-1}-1)/2]}\ mod\ m$\\
        &$5:\ \textbf{for}\ i \leftarrow n-2\ downto\ 0$\\
        &$6:\ \quad x \leftarrow x^{2^{L(w_{i})}} mod\ m$\\
        &$7:\ \quad \textbf{if}\ w_{i} \neq 0\ \textbf{then}$\\
        &$8:\ \quad \quad x \leftarrow x \cdot \textcolor{blue}{g[(w_{i}-1)/2]}\ mod\ m$\\
        &$9:\ \textbf{return}\ x$
    \end{tabular}
    \caption{Sliding-window Exp.}
    \label{fig:channel2}
    \end{subfigure}
\end{minipage} 
\caption{Cache Side-channel Examples.}
\label{fig:channels}
\vspace{-10pt}
\end{figure}



\subsection{Cache Side Channel Mitigation}
\label{subsec:prel-defense}

\cite{lou2021survey} surveys software-level countermeasures of cache side
channels. Overall, two code patterns can remove secret-dependent cache access
patterns: \textit{AlwaysAccess-BitwiseSelect} permits programs to access
secret-dependent data within each loop iteration in a constant manner, while
deciding whether or not to accept it via bitwise operations. Moreover, if the
calculation is inexpensive and free of secret-dependent branches,
\textit{On-the-fly Calculation} avoids using lookup tables, which eliminates
leakage shown in \F~\ref{fig:channel2}.
Similarly, to remove secret-dependent branches,
\textit{AlwaysExecute-ConditionalSelect} enables covering all branches
regardless of the \texttt{if} conditions. 
\textit{AlwaysExecute-BitwiseSelect} eliminates secret-dependent branches by
selecting correct results through bitwise operations.

The aforementioned code patterns can frequently introduce high overhead. They
are thus less frequently used to only secure several core code fragments, which
may miss subtle usage of secrets~\cite{doychev2017rigorous, wang2017cached}. 
\textit{Blinding} introduces extra randomness in crypto computations
to obscure the inference of secrets. Depending on the blinding target,
there are two distinct usages of blinding masks.

\noindent \textbf{Key Blinding.}~With this scheme enabled, the attacker obtains
blinded secrets without knowing the blinding mask $r$. As $r$ is randomly
generated before each cipher process, attacker cannot exploit the cryptosystem.
For example, exponent blinding in RSA adds a random multiple of Euler's $\phi$
function, i.e., $r \cdot \phi(n)$, to the secret exponent. Then, RSA decryption
performs $c^{d+r \cdot \phi(n)}\ mod\ n$, which equals
$c^{d}\ mod\ n$. Though some known attacks~\cite{schindler2015exclusive}
exploit this scheme, the exponent blinding still impedes the attacker at large.

\noindent \textbf{Plaintext/Ciphertext Blinding.}~Blinding can also be applied
to plaintext/ciphertext. For instance, when enforcing blinding, RSA converts
the ciphertext $m$ into $m \cdot r^{e}$, where $r$ is the random factor. The
original result $m^{d}\ mod\ n$ can be obtained by multiplying the new result
$(m \cdot r^{e})^{d}\ mod\ n$ by $r^{-1}$ due to $r^{ed} \cdot r^{-1} mod\ n
\equiv 1\ mod\ n$. The plaintext/ciphertext blinding defeats known-input
attacks that leverage timing side channels.

Blinding can usually provide more comprehensive protection as once
key/ciphertext is blinded, all their follow-up usages and their (subtle)
influence on other variables should be protected. \rev{However, their effectiveness in
mitigating cache side channels are not yet comprehensively analyzed, given the
difficulty of modeling them automatically in previous methods (noted in
\textbf{Challenge 3} in \S~\ref{sec:introduction}).} 

%% file: 3_overview.tex
\section{Research Overview}
\label{sec:overview}

\begin{figure*}[!htbp]
\centering
\includegraphics[width=0.90\textwidth]{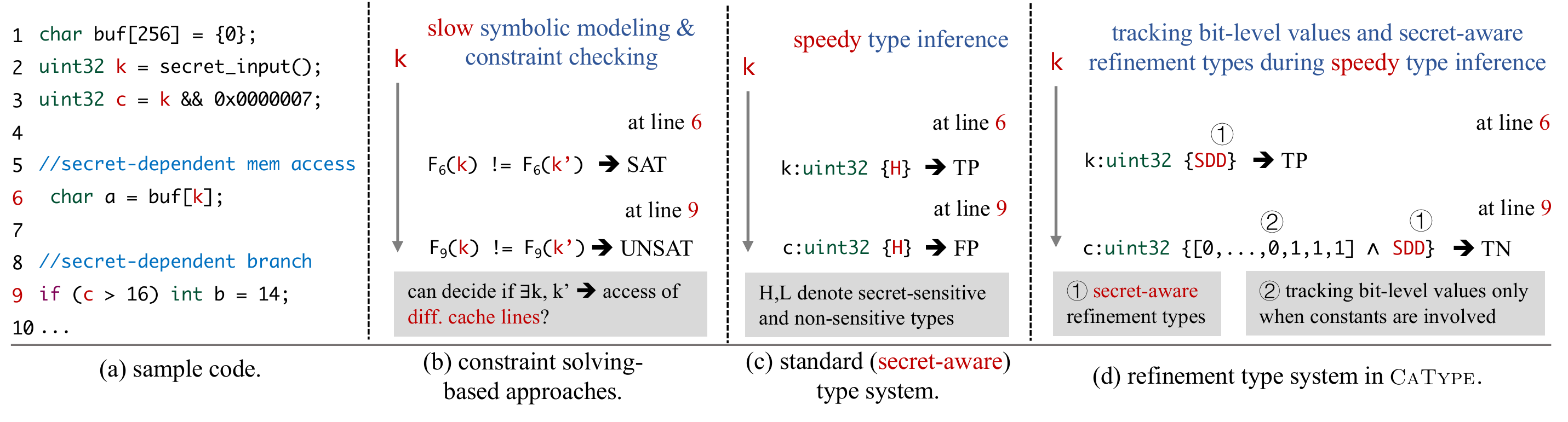}
\caption{Comparison of constraint solving-based techniques (b), type
inference-based approach (c), and \tool\ (d). TP, FP, and TN denotes true
positive, false positive, and true negative, respectively.}
\label{fig:motivation}
\vspace{-5pt}
\end{figure*}

\subsection{Assumptions}
\label{subsec:assumption}

\noindent \textbf{Threat Model.}~\rev{\tool\ follows an identical threat model
as most current cache side channel detectors~\cite{wang2017cached,
wang2019identifying, jan2018microwalk, brotzman2019casym, bao2021abacus}. We
assume that an adversary shares the same hardware platform as the victim, a
typical and practical assumption in cloud computing systems. Thus, while the
adversary cannot directly monitor the victim's memory accesses, he can probe the
shared cache states to determine if certain cache lines have been visited by the
victim software. This threat model covers the majority of cache side channel
attacks in the literature. For example, adversaries infer cache accesses by
measuring the latency of the victim program in $\textsc{EVICT-TIME}$
attack~\cite{osvik2006cache}, or the latency of the attacker program in
$\textsc{PRIME-PROBE}$~\cite{percival2005cache, osvik2006cache, liu2015last},
$\textsc{FLUSH-RELOAD}$~\cite{yarom2014flush}, and $\textsc{FLUSH-FLUSH}$
attacks~\cite{gruss2016flush}.}

\rev{Existing works~\cite{doychev2013cacheaudit,brotzman2019casym}
commonly refer to the attackers in our threat model as ``trace-based attackers''
since they are able to probe the cache state after the execution of each program
statement in the victim software.}
It is also worth noting that the attackers can distinguish cache layouts of
instructions inside the program branches of shared libraries. This is due to the
fact that modern OSes adopt aggressive memory deduplication techniques, allowing
shared libraries to be mapped to copy-on-write pages. As a result, the
probing granularity of attackers is precisely reduced to cache lines.

\noindent \textbf{Main Audience.}~\rev{Consistent with previous
works~\cite{wang2017cached, wang2019identifying, jan2018microwalk,
bao2021abacus, doychev2013cacheaudit, daniel2020binsec, brotzman2019casym,
weiser2018data, chattopadhyay2019quantifying, sung2018canal}, \tool\ is
primarily designed for crypto software developers who have sufficient knowledge
about their own software. Before release, \tool\ serves as a ``vulnerability
debugger'' for the developers to detect attack vectors in their software. \tool\
provides fully automated and speedy analysis to flag program points that leak
secrets via cache side channels. Developers can accordingly patch \tool's
findings to mitigate leakage. Nevertheless, we clarify that \tool\ is
\textit{not} an attack tool; the exploitability of its findings (e.g., whether
RSA private keys can be reconstructed via \tool's findings) is beyond the scope
of this paper.}

\subsection{Methodology Overview}
\label{subsec:motivation}

\rev{This section illustrates the high-level methodology overview and compares with
existing efforts in \F~\ref{fig:motivation}. Recall that we have introduced two
typical cache side channel patterns in \S~\ref{sec:background}: SDMA and SDBC.
\F~\ref{fig:motivation}(a) presents a sample code that is vulnerable to SDMA
(line 6) whereas the condition at line 9 is \textit{not} vulnerable to SDBC,
given that the \texttt{else} branch will always be executed.}

\parh{Symbolic Execution-Based Approaches.}~\rev{De facto side channel detectors
perform heavyweight symbolic execution, where program (secret-related) data
facts are modeled using symbolic formulas. Then, at each memory access and
branch condition, they check if different secrets can lead to the access of
different cache lines using constraint solving. For instance, let symbol $k$
represent the secret read in line 2 of \F~\ref{fig:motivation}(a), existing side
channel
detectors~\cite{wang2017cached,wang2019identifying,bao2021abacus,brotzman2019casym}
primarily check the following constraint to decide SDMA/SDBC:}

\vspace{-3pt}
\begin{equation}
\label{condition1}
\exists k \neq k', \ F(k) \neq F(k') 
\end{equation}

\noindent \rev{where $F$ denotes the memory access constraint formed at line 6, or
branch condition constraint formed at line 9. The symbolic engine forms $F(k) =
b + k \times 4$ at line 6, where $b$ is the base address of
\texttt{buf}. The satisfiability (SAT) of Constraint~\ref{condition1}
checks the existence of two secrets that lead to the access of different
cache lines, such that certain amount of secrets will be leaked to
the attacker. Moreover, the symbolic engine will track computations using
symbolic formulas, and at line 9, the constraint solver yields unsatisfiable
(UNSAT) for Constraint~\ref{condition1}, thereby proving the safety of line 9.}

\rev{The primary obscurity of such detectors is \textit{scalability}. Overall,
existing symbolic execution (or abstract interpretation)-based side channel
detectors need to maintain complex symbolic states for each program statement to
encode program semantics. As symbolic execution continues, the symbolic constraints (encoding program states) will steadily accumulate and grow in size,
filling a vast amount of memory. Even worse, existing tools need to perform
constraint solving for each suspicious memory access and conditional branch
instruction, and constraint solving is generally slow. With this regard, we
notice that existing static analysis tools are often limited to analyzing small
programs, or fail to consider the effect of side channel mitigation techniques
like blinding.}

\parh{Conventional Type-Based Analysis.}~\rev{\S~\ref{subsec:refinement} has
introduced basic mechanisms of type systems and the extensions to track high/low
secret-sensitive data with type annotations $H$ and $L$. As illustrated in
\F~\ref{fig:motivation}(c), performing type inference can easily establish that
the types of \texttt{k} and \texttt{c} (in lines 6 and 9, respectively) are
\texttt{uint32}. Moreover, by assigning a high security sensitivity type $H$ to
\texttt{k} at line 2, the type system identifies two usage of sensitive data at
line 6 and line 9. These two statements are deemed as ``vulnerable'', leading to
secret-dependent memory access and branch condition. Nevertheless, we underlie
that while the statement at line 6 is a true positive (TP) finding, statement at
line 9 is a false positive (FP), as \texttt{c} can never exceed 7 (see line 3 in
\F~\ref{fig:motivation}(a)). Overall, conventional type-based analysis delivers
speedy tracking of (secret-related) data through type annotations. They,
however, lack of tracking values and are less expressive than constraint
solving-based methods. Indeed, \S~\ref{sec:discussion} compares taint analysis,
\textit{conceptually similar} to type systems enforcing information-flow
security (e.g.,~\cite{DBLP:journals/jsac/SabelfeldM03}), with refinement type
system implemented in \tool. We show that taint analysis yields considerably
more false positives than \tool.}

\parh{Refinement Type System in \tool.}~\rev{Recall the refinement type of a
variable $x$ can be expressed as $x:T\{v:P\}$ (\S~\ref{subsec:refinement}),
where $T$ and $P$ are basic types and predicates, respectively.
\F~\ref{fig:motivation}(d) illustrates the usage of the refinement type system in
\tool, where the refinement formalizes the concerned
(secret-related) program properties as predicates. In particular, we use type
$\sdd$ to denote secret-dependent values, and the refinement type system
infers that in line 6, \texttt{k} is of type $\text{uint32} \{v: \sdd\}$,
revealing a potential SDMA case. Similarly, the refinement type of \texttt{c} in
line 9 also has type $\sdd$, revealing a potential SDBC case (which is
\textit{not} vulnerable; see below for clarification). \tool\ defines in total
five predicates, systematically considering secret-dependent,
secret-independent, as well as blinding operations. In this way, \tool\ can benefit
from refinement type techniques to keep track of secret propagations and
identify SDMA/SDBC in a speedy manner while correctly considering randomization
mechanisms like blinding (see \textbf{Blinding} later this section for further
discussion).}

\rev{Moreover, \tool\ explores an important improvement, by tracking bit values
directly in refinement types, in the form of value predicates. A value predicate
is defined as $v = b$, where $b$ is either $0$ or $1$. 
\tool\ is carefully designed to deliver a ``mild tracking'' of bit-level values.
That is, only the refinement types of constants are initialized to comprise
bit-level predicates. Then, \tool\ tracks the bit-level predicates via type
inference in a correct yet conservative manner. For instance, when a constant,
0x0000007, is used as the mask over the secret (line 3), the type of the output means that it is a bitvector
with all secret bits (except the three least significant bits) set to 0.
Note that
value predicates in refinement types can be absent, indicating that the precise
bit-level values are unknown.}

\rev{By tracking of bit values from constants, \tool\ can exclude the
majority, if not all, cases where different secret values at a suspicious
SDMA/SDBC case result in visiting the \textit{same} cache line (i.e., a safe
program site). For instance, when \texttt{k} is masked by 0x0000007 before
being used in the \texttt{if} condition at line 9 of \F~\ref{fig:motivation}(a),
the refinement type of \texttt{c} has all bits set to 0 except the lowest three
bits, and \tool\ can simply decide that the branch condition will always be
evaluated as ``false'' with an arithmetic comparison over two bitvectors.
Therefore, when analyzing the statement at line 9 of \F~\ref{fig:motivation}(a),
\tool\ yields a true negative (TN) finding, as shown in
\F~\ref{fig:motivation}(d). Overall, we view that the refinement type system
designed in \tool\ manifests comparable capability with constraint solving-based
methods to analyze cache side channels. Moreover, \tool\ avoids the use of
constraint solving, and is therefore dramatically faster; see
\T~\ref{tab:performanceoverview} in \S~\ref{subsec:results}.}

\parh{Potential False Positives.}~\rev{We clarify that the refinement type
system in \tool\ may not always know the precise bit values: the absence of
value predicates means the value could be $0$ or $1$. Overall, \tool\ tracks the
bit values introduced by constants using refinement types at ``its best
effort''. Thus, we may encounter false positives, e.g., due to constants that are
however not tracked by \tool. Nevertheless, cache side channels are rare in
practice, and we confirm that all findings of \tool\ over production
cryptosystems are true positives. Also, the refinement type system is sound
without introducing false negatives, as benchmarked in \S~\ref{sec:discussion}.}

\parh{Blinding.}~\rev{As introduced in \S~\ref{subsec:prel-defense}, modern
cryptosystems use randomness mechanisms like blinding to impede side channels.
To capture the security property of blinding, our refinement type system
facilitates a smooth and accurate modeling of blinding, by adding specific
predicates in type refinement to denote uniformly random data (i.e., the
blinding mask). 
We also define type inference rules and propagation rules for blinding involved computations, so that we can capture sufficient
information used to infer potential leaks. For example,
uniformly random factors can perfectly mask the result through logic xor
operation, eliminating the effects of a secret if it is a source operand. See
details in \S~\ref{typefunction} and \S~\ref{typesystem}.}

\rev{In contrast, adding support for blinding presumably increases the search
space of constraint solving-based methods to a great extent. Consequently, finding a
SAT solution for Constraint~\ref{condition1} is highly expensive, especially
when both secrets and blinding masks are present. Though an ``optimal
solution'' is not yet clear, inspired by relevant research in perfect masking
analysis~\cite{eldib2014qms, eldib2014smt, eldib2014formal}, we expect to fix 
two different secrets $k,k'$ and then iterate the quantifiers of all involved masks 
$r_{1}, \dots, r_{n}$ to count the ranges under $k,k'$. This process may take 
a dramatically longer time or timeout.}


\begin{figure}[htbp]
\centering
\includegraphics[width=0.47\textwidth]{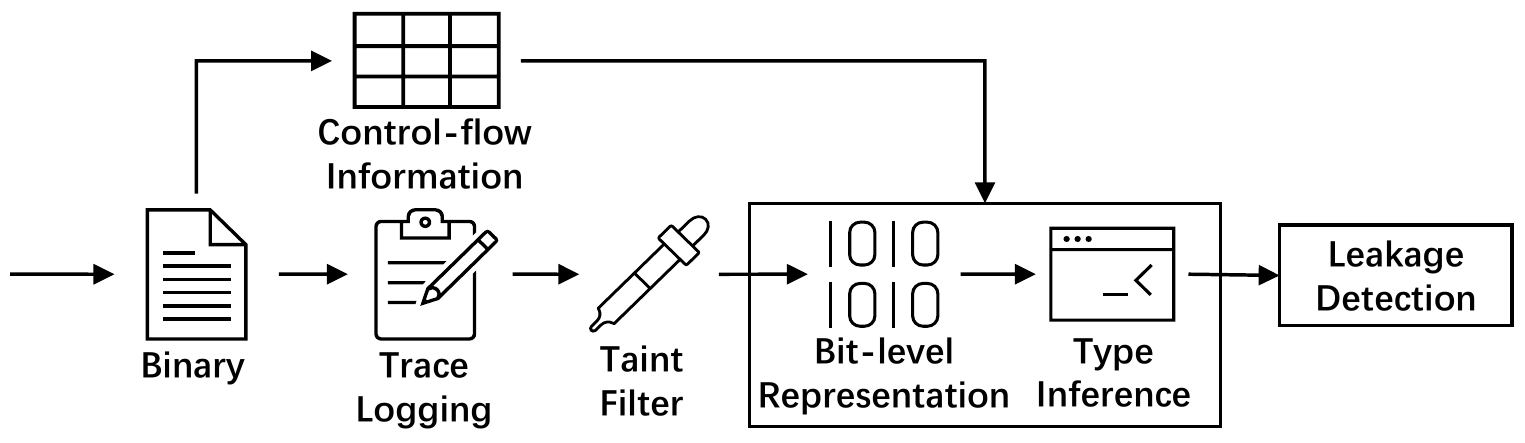}
\caption{Workflow of \tool.}
\label{fig:architecture}
\vspace{-5pt}
\end{figure}

%% file: 4_design.tex
\section{Design}
\label{sec:design}

\parh{Overview.}~\F~\ref{fig:architecture} depicts the workflow of \tool. Given
the crypto software in executable format, we first run the executable using
Intel Pin~\cite{luk2005pin} to perform concerned crypto computation (e.g., RSA
decryption) and log an execution trace. Then, we require users of \tool\ to mark
the program secrets and random factors on the execution trace, and perform taint
analysis by tainting those secrets/randomness and extract a tainted sub-trace
depicting how tainted variables are propagated and used. Meanwhile, we also
disassemble the executable and extract control flow information into a lookup
table from the disassembled assembly code, which will be used later in checking
SDBC (see \S~\ref{subsec:detection}).

\tool\ then performs type inference over the tainted sub-trace, by first
annotating variables with bit-level types of initialized refinements
(\S~\ref{bitlevel}). It tracks the propagation and usage of
secure-sensitive values in refined types during type inference
(\S~\ref{typefunction} and \S~\ref{typesystem}). When encountering memory accesses
or branch conditions, \tool\ uses the refined types of involved variables to
check if SDBC/SDMA exists (\S~\ref{subsec:detection}). Once a side channel flaw
is discovered, it reports the detected
instruction's address to users for confirmation, debugging, and patching. We
now discuss each step in detail.

\noindent \textbf{Design Consideration: Binary vs. Source.}~\rev{\tool\ is
designed to directly analyze x86 binary code compiled from crypto software.
Thus, the refinement type system is defined over x86 assembly code, and \tool's
analysis depends on the specific memory layout. Overall, side channels are
sensitive to the low-level architecture and system details. We clarify that
prior works in this field are consistently analyzing software in executable
format. This enables the analysis of legacy code and third-party libraries
without accessing source code. More importantly, by analyzing low-level assembly
instructions, it is possible to take into account low-level details, such as
memory allocation. Recent works~\cite{simon2018you} have shown that compiler
optimizations could introduce extra side channel opportunities that are not
visible at the high-level code representation level.}

\noindent \textbf{Design Consideration: Information Flow Tracking.}~\rev{When
illustrating cache side channels in \F~\ref{fig:channels} and
\F~\ref{fig:motivation}, we depict how the use of secrets result in side
channels. Nevertheless, in addition to side channels induced via the
\textit{direct} usage of secrets, it is crucial to treat data derived from the
secrets as ``sensitive''. \textit{\tool\ tracks both explicit and implicit
information flows propagated from secrets.} When a variable $x$ is of $\sdd$
type, and the data is loaded from memory address formed by $x$, the  
destination variable has type $\sdd$. Similarly, when $x$ is used to form branch
conditions, the result type is $\sdd$ as well. By modeling information flows,
\tool\ comprehensively uncovers attack surface of cryptosystems.}

\subsection{Bit-level Representation and Types}
\label{bitlevel}

We first clarify that in analyzing x86 assembly code, registers, CPU flags, and
memory cells are all considered as \textit{variables} in \tool. We use bit-level
representation for variables encountered on the execution trace, allowing us to
track variables with fine-grained precision. Considering the instruction syntax
in \F~\ref{fig:syntax-new}, where an expression $e$ can be a constant bit $b$, a
variable $x$, a constant bitvector $[b, \cdots, b]$, or computations over
expressions. Concatenation $e_1 \sharp\ e_2$ uses $e_1$ and $e_2$ to form the
highest and lowest several bits, respectively. Extracting several bits from the
designated position of a bitvector expression produces a fragment, dubbed as
$[n_1:n_2]/e$. Other operations include negation ($\neg$), arithmetic and logic
operations ($\bowtie$) over two expressions, and the conditional expression with
three operands (the syntax mimics conditional selection in the C language). 
A statement $s$ is an assignment, a memory load/store, or a sequence of 
statements. We clarify that execution trace forms a typical straight-line code of instructions, omitting branch merges.

\parh{Types and Hierarchy.}~As introduced in \S~\ref{subsec:motivation}, a type
$\rho$ has the form of $\{v: T \mid P\}$, where $T$ is a basic type and
predicate $P$ is the refinement. We define basic type $T$ as primitive types of
bit representations, i.e., one bit $\text{B}$ or a bitvector of $n$ bits
$\bv{n}$. A refinement type $P$ is either a security type predicate $\tau$ or a
conjunction with a value predicate. A security type predicate $\tau$ can be any
of the five types, i.e., $\sdd$, $\rad$, $\sid$, \rev{$\ukd$} and $\cst$, denoting
\textit{secret-dependent}, \textit{uniformly random},
\textit{secret-independent}, \rev{\textit{weakly random}}, and \textit{constant}
values. A value predicate is termed as $v = b$ (where $b$ is 1 or 0), meaning
that $v$ has value $b$. The expression typing judgment, $\Gamma \vdash e: \rho$,
states that expression $e$ has type $\rho$, where $\Gamma$ is the typing
environment mapping from variables to types.

The hierarchy of security types $\tau$ is $\cst \subtype \ura \subtype
\wra \subtype \sid \subtype \sdd$. \rev{We clarify that among the five refined types,
only $\sdd$ is related to secrets. 
We use \wra\ to denote a data of weakly random distribution, meaning it is not
uniformly random (in other words, not perfect and secure blinding). \ura\ means  
uniformly random data, representing perfect and secure masking.} The join
operator $\sqcup$ takes the least upper bound of two types; for instance, $\sid
\sqcup \sdd = \sdd$, as $\sdd$ sits higher in the hierarchy.

\parh{Types Annotation.}~Before launching type inference, we first annotate variables with security
types. Secrets, random factors, and constants are marked as $\sdd$, $\rad$, and
$\cst$, respectively. We mark other variables using $\sid$, and type $\ukd$ may
be generated during type inference. Given that we perform bit-level type
annotation and inference, if variable $x$ hosts a 32-bit secret, it is annotated
as $\{v: \bv{32} \mid v: \sdd \}$. This vector type implies that each bit in the
vector has type $\sdd$, i.e., $\forall b_{i} \in x.\ b_{i}: \{v: B \mid v: \sdd
\}$. 
For constants, we also explicitly annotate each bit (whether it equals 0
or 1) in the value predicate. Thus, each bit of a constant $c$ is in the form of
$b_{i} \in c.\ b_{i}: \{v:B \mid v = b \land v: \cst \}$, where $b$ is $0$ or
$1$, depending on the value of $c$. Recall as noted in
\S~\ref{subsec:motivation}, our refinement type-based inference conducts a
``best-effort'' tracking of bit-level values derived from constants. The
bit-level tracking updates value predicates during type inference. Nevertheless,
when a bit value becomes unknown (could be either 0 or 1), we conservatively
omit its value predicate and only retain the security type predicate. 

\begin{figure}[t]
\centering
\footnotesize
\[
\begin{array}{r l l l}
    \text{Expr} & e & ::= & b \mid x \mid [b, \cdots, b] \mid \neg e \mid e_1 \bowtie e_2 \\
    & & \mid & e\ ?\ e_1:e_2 \mid e_1\ \sharp\ e_2 \mid [n_1:n_2]/e \\
    \text{Stmt} & s & ::= & x \leftarrow e \mid x \leftarrow e_1[e_2] \mid e_1[e_2] \leftarrow x \mid s_1;s_2 \\
    \text{Basic Types} & \text{T} & ::= & \text{B} \mid \bv{n} \\
    \text{Security Types} & \tau & ::= & \sdd \mid \rad \mid \sid \mid \ukd \mid \cst\\
    \text{Refinements} & \text{P} & ::= & v: \tau \mid v = b \land v: \tau \\
    \text{Type} & \rho & ::= & \{v: \text{T} \mid \text{P}\} \\
    \text{Type Env} & \Gamma & ::= & \emptyset \mid \Gamma, x: \rho \\
\end{array}
\]
\caption{Syntax of bit-level representation.}
\label{fig:syntax-new}
\end{figure}



\subsection{Type Inference for Bitvectors}
\label{typefunction}

\begin{figure}
\centering
\footnotesize
\begin{gather*}
\lVert [b_{n-1}, \cdots, b_0] \rVert_{t}=
\begin{cases}
\sdd & \text{$\exists b_{i}.\ b_{i}: \{v:B \mid v: \sdd \}$}\\
\rad & \text{$(\nexists b_{i}.\ b_{i}: \{v:B \mid v: \sdd \})\ \wedge $}\\
    & \text{$(\exists b_{i}.\ b_{i}: \{v:B \mid v: \rad \})$}\\
\sid & \text{$(\nexists b_{i}.\ b_{i}: \{v:B \mid v: \sdd \})\ \wedge $}\\
    & \text{$(\nexists b_{i}.\ b_{i}: \{v:B \mid v: \rad \})\ \wedge $}\\
    & \text{$(\exists b_{i}.\ b_{i}: \{v:B \mid v: \sid \})$}\\
\ukd & \text{$(\nexists b_{i}.\ b_{i}: \{v:B \mid v: \sdd \})\ \wedge $}\\
    & \text{$(\nexists b_{i}.\ b_{i}: \{v:B \mid v: \rad \})\ \wedge $}\\
    & \text{$(\nexists b_{i}.\ b_{i}: \{v:B \mid v: \sid \})\ \wedge $}\\
    & \text{$(\exists b_{i}.\ b_{i}: \{v:B \mid v: \ukd \})$}\\
\cst & \text{$\forall b_{i}.\ b_{i}: \{v:B \mid v: \cst \}$}\\
\end{cases}\nonumber
\end{gather*}
\caption{Type propagation from single-bit to bitvector.}
\label{fig:distribution}
\end{figure}

Different bits in a bitvector may have varying security types. Consider register
\texttt{eax}, which stores a 32-bit data, where the upper 16 bits are
$\rad$ and the lower 16 bits are $\sid$. Intuitively, the bitvector's type can
be inferred by simply taking the least upper bound of all the constituent
bits' types, i.e., $\sid$ in this case. However, the high 16 bits are $\rad$, meaning that each bit has equal possibility of being 0 or 1. Thus,
the intuitive approach would lose the
information of randomness, leading to inaccuracy in subsequent analyses.

To precisely track bit-level security propagation, we define function $\lVert x
\rVert_{t}$ in \F~\ref{fig:distribution} to infer a bitvector's type from the
types of its constituent bits based on a notion of structural priority.
We give type $\sdd$ the highest priority, meaning that a bitvector is of type
$\sdd$ if it contains at least one bit of type $\sdd$. In the absence of $\sdd$
type, type $\rad$ is structurally preceding, i.e., if there is a bit in a vector
whose type is $\rad$, then the vector itself is $\rad$. As seen in
\F~\ref{fig:distribution}, $\sid$ is structurally superior to $\ukd$ and
$\cst$, whereas $\ukd$ is structurally superior to $\cst$. 

\rev{From a holistic view, sensitive data (specified in refinements) are
``propagated'' from single-bit to whole bitvector following type rules in
\F~\ref{fig:distribution}.
Therefore, information flow analysis is performed here to determine how sensitive
data are propagated and influence program execution. To clarify, in addition to
type rules, \tool\ also conducts taint analysis over the Pin-logged trace and
collects a list of tainted instructions. This is a classic optimization to reduce
trace length, also adopted in previous
works~\cite{wang2017cached,wang2019identifying,bao2021abacus}. Our type
inference is performed on the tainted trace, as illustrated in
\F~\ref{fig:architecture}.}



\subsection{Type Inference Rules}
\label{typesystem}

\begin{figure*}[htbp]
\footnotesize
\begin{mathpar}
\inferrule[Conj$\&$Disj.I]{\Gamma \vdash e_1: \{v:B \mid v: \tau_1 \} \\ \Gamma \vdash e_2 : \{v: B \mid v: \tau_2\} \\\\ \tau_1 \neq \cst \\ \tau_2 \neq \cst \\ \neg (\tau_1 = \rad \wedge \tau_2 = \rad) \\ \bowtie\ \in \{ \wedge,\vee \}}
{\Gamma \vdash e_1 \bowtie e_2 : \{v:B \mid v: \tau_1 \sqcup \tau_2\}}
\and
\inferrule[Conj$\&$Disj.II]{\Gamma \vdash e_1: \{v: B \mid v: \rad\} \\ \Gamma \vdash e_2 : \{v: B \mid v: \rad\} \\ \bowtie\ \in \{ \wedge,\vee \}}
{\Gamma \vdash e_1 \bowtie e_2 : \{v: B \mid v: \ukd \}}
\and
\inferrule[XOR.I]{\Gamma \vdash e_1: \{v: B \mid v: \tau_1\} \\ \Gamma \vdash e_2: \{v:B \mid v: \tau_2 \} \\\\ \tau_1 \neq \rad \\ \tau_2 \neq \rad \\ \neg(\tau_1 = \cst \land \tau_2 = \cst)}
{\Gamma \vdash e_1 \oplus e_2: \{v:B \mid v: \tau_1 \sqcup \tau_2\}}
\and
\inferrule[XOR.II]{\Gamma \vdash e_1: \{v: B \mid v: \rad \} \\ \Gamma \vdash e_2: \{v: B \mid v: \tau \}}
{\Gamma \vdash e_1 \oplus e_2: \{v: B \mid v: \rad \}}
\and
\inferrule[Neg.I]{\Gamma \vdash e: \{v:B \mid v: \tau \}}
{\Gamma \vdash \neg e: \{v: B \mid v: \tau \}}
\end{mathpar}
\caption{Selected one bit $\text{B}$ type rules for logical operations. See
Appendix~\ref{appdix:rules} for the complete list of rules.}
\label{fig:bitrule-new}
\end{figure*}

\begin{figure*}[htbp]
\footnotesize
\begin{mathpar}
\inferrule[Concat.I]{\Gamma \vdash e_1 : \{v:\bv{n_1} \mid v: \tau_1\} \\ \tau_1 \neq \rad \\\\ \Gamma \vdash e_2 : \{ v:\bv{n_2} \mid v: \tau_2\} \\ \tau_2 \neq \rad \\ }
{\Gamma \vdash e_1\ \sharp\ e_2 : \{v: \bv{n_1+n_2} \mid v: \tau_1 \sqcup \tau_2 \}}
\and
\inferrule[Concat.II-1]
{\Gamma \vdash e_1 : \{v:\bv{n_1} \mid v: \rad\} \\\\ \Gamma \vdash e_2 : \{v: \bv{n_2} \mid v: \tau_2\} \\ \tau_2 \neq \sdd }
{\Gamma \vdash e_1\ \sharp\ e_2 : \{v: \bv{n_1+n_2} \mid v: \rad \}}
\and
\inferrule[Concat.II-2]{\Gamma \vdash e_1 : \{v:\bv{n_1} \mid v: \rad\} \\\\ \Gamma \vdash e_2 : \{v: \bv{n_2} \mid v: \sdd\} }
{\Gamma \vdash e_1 \sharp e_2 : \{v: \bv{n_1+n_2} \mid v: \sdd \}}
\and
\inferrule[Extraction]
{\Gamma \vdash e: \{v: \bv{n} \mid v: \tau_e \} \\\\ m_1 \leq m_2 \\ \lVert \lbrack m_1:m_2 \rbrack / e \rVert_{t} = \tau}
{\Gamma \vdash \lbrack m_1:m_2 \rbrack / e: \{v: \bv{m_2-m_1+1} \mid v: \tau \}}
\and
\inferrule[Logic.I]{ \Gamma \vdash e_1: \{v:\bv{n} \mid v: \tau_1\} \\ \Gamma \vdash e_2: \{v:\bv{n} \mid v: \tau_2 \} \\\\ \bowtie \in \{ \land , \lor, \oplus \} \\ \lVert e_1 \bowtie e_2 \rVert_{t} = \tau }
{\Gamma \vdash e_1 \bowtie e_2 : \{v : \bv{n} \mid v: \tau\}}
\and
\inferrule[Logic.II]{\Gamma \vdash e: \{v: \bv{n} \mid v: \tau\}}
{\Gamma \vdash \neg e : \{v: \bv{n} \mid v: \tau\}}
\and
\inferrule[Arith.I]{\Gamma \vdash e_1: \{v: \bv{n} \mid v: \tau_1\} \\ \Gamma \vdash e_2: \{v: \bv{n} \mid v: \tau_2\} \\\\ \tau_1 \neq \rad \\ \tau_2 \neq \rad \\ \neg(\tau_1 = \cst \land \tau_2 = \cst) \\ \bowtie \in \{ +, -, \times, \div \}}
{\Gamma \vdash e_1 \bowtie e_2 : \{v: \bv{n} \mid v: \tau_1 \sqcup \tau_2\}}
\and
\inferrule[Arith.II-1]{\Gamma \vdash e_1: \{v: \bv{n} \mid v: \rad\} \\ \Gamma \vdash e_2: \{v: \bv{n} \mid v: \tau_2 \} \\\\ \tau_2 \neq \sdd \\ \bowtie \in \{ +, -, \times, \div \}}
{\Gamma \vdash e_1 \bowtie e_2 : \{v: \bv{n} \mid v: \rad \}}
\and
\inferrule[Arith.II-2]{\Gamma \vdash e_1: \{v: \bv{n} \mid v: \rad\} \\\\ \Gamma \vdash e_2: \{v: \bv{n} \mid v: \sdd \} \\ \bowtie \in \{ +, -, \times, \div \}}
{\Gamma \vdash  e_1 \bowtie e_2 : \{v: \bv{n} \mid  v: \sdd\}}
\and
\inferrule[Comp]{\Gamma \vdash e_1: \{v: \bv{n} \mid v: \tau_1 \} \\ \Gamma \vdash e_2: \{v: \bv{n} \mid v: \tau_2 \} \\\\ \neg(\tau_1 = \cst \land \tau_2 = \cst) \\ \bowtie \in \{ <, \leq, >, \geq, =, \neq \}}
{\Gamma \vdash e_1 \bowtie e_2: \{v:\bv{1} \mid  v: \tau_1 \sqcup \tau_2\}}
\and
\inferrule[Cond.I]{ \Gamma \vdash e:\{v: \bv{1} \mid v: \sdd\} \\ \Gamma \vdash e_1: \{v: \bv{n} \mid v: \tau_1\} \\\\ \Gamma \vdash e_2: \{v: \bv{n} \mid v: \tau_2\}}
{\Gamma \vdash e\ ?\ e_1: e_2: \{v: \bv{n} \mid v: \sdd\}}
\and
\inferrule[Cond.II]{ \Gamma \vdash e:\{v: \bv{1} \mid v: \tau\} \\ \tau \neq \sdd \\ \Gamma \vdash e_1: \{v: \bv{n} \mid v: \tau_1\} \\\\ \Gamma \vdash e_2: \{v: \bv{n} \mid v: \tau_2\} \\ \neg(\tau_1 = \cst \land \tau_2 = \cst)}
{\Gamma \vdash e\ ?\ e_1: e_2: \{v: \bv{n} \mid v: \tau_1 \sqcup \tau_2\}}
\end{mathpar}
\caption{Type rules for expressions involving bitvector $\bv{n}$.}
\label{fig:typerule-new}
\end{figure*}

\tool\ implements a comprehensive set of type inference rules over each
encountered x86 assembly instruction to track the propagation of
secure-sensitive types and check cache side channels. 

\noindent \textbf{Type Rules for One Bit Logical
Operations.}~\F~\ref{fig:bitrule-new} presents a representative list of type
rules for one bit logical operations. First, type rules that involve
$\cst$ type are designed to propagate $\cst$ in a straightforward way; see
Appendix~\ref{appdix:rules} for other involved rules. Rule
\textsc{Conj$\&$Disj.I} states that if two operands are not both $\cst$ or
$\rad$, then the result type is the least upper bound of the two operands'
types, which enables the tracking of secure-sensitive values in types. Rule
\textsc{Conj$\&$Disj.II} handles the circumstance in which both operands are
$\rad$. Since the value of the result is no longer distributed uniform-randomly
under logic \textit{AND} and \textit{OR}, the result type is lifted on the type
hierarchy to $\ukd$.

Rule \textsc{XOR.I} is similar to rule \textsc{Conj$\&$ Disj.I}, where the
result type is the least upper bound of the two operands' types, provided that
neither bit expression is $\rad$ or $\cst$ simultaneously. Rule \textsc{XOR.II}
states that if one of the operands is of type $\rad$, the result type is $\rad$.
This refers to the fact that random factors can uniformly blind the results
through exclusive or ($\oplus$) operations.
Rule \textsc{Neg.I} keeps security types unchanged in front of the negation
operation.

\noindent \textbf{Type Rules for Bitvector
Operations.}~\F~\ref{fig:typerule-new} depicts the type rules for operations
with bitvectors $\bv{n}$. There are three rules applicable to concatenation
expressions. Rules \textsc{Concat.I} states that the resultant's type takes the
least upper bound of the two vectors' type, if both vectors are not $\rad$. Rule
\textsc{Concat.II-1} states that type $\rad$ is structurally prior to other
secret-free types, and \textsc{Concat.II-2} specifies that a bitvector exhibits
$\sdd$ type if at least one bit in expression $e_2$ is $\sdd$. Rule
\textsc{Extraction} is a well-demonstrated example that leverages function
$\lVert x \rVert_{t}$ to determine the refined type of the segment extracted
from the source operand. Note that shift operations do not have their own rules
as they can be implemented by combining concatenation and extraction operations.
Rule \textsc{Logic.I} infers a vector type from the types of its constituent
bits, i.e., the type of the result is inferred by applying the structural
priority defined in \F~\ref{fig:distribution}. Rule \textsc{Logic.II} is similar
to Rule \textsc{Neg.I}.

For the arithmetic operations of two bitvectors, one difference lies in
performing the calculation at the whole bitvector level as opposite to each bit.
Specifically, we determine the security type of the result, and
propagate it to each bit; this offers a sound estimation of each bit's
security type. 
Similar to \textsc{Concat} rules, \textsc{Arith} rules conform to the security
type propagation in bitvector structures.

As specified in x86 assembly code, the comparison operation only produces
one-bit bitvector $\bv{1}$ to the result (i.e., the affected CPU flags). Rule
\textsc{Comp} specifies that the resultant's type is the least upper bound of
the two operands' types. We omit the case where two operands are both $\cst$ as
it is straightforward. The last two rules are designed for conditional
expressions. We specify two rules according to whether the condition expression
$e$ is related to the secret. Rule \textsc{Cond.I} states that if the refined
security type of the condition expression $e$ is \sdd, the result type is \sdd\
regardless of the type of two branch expressions. \rev{We clarify that this rule
allows \tool\ to keep track of implicit information flow propagated from
secret-dependent branch conditions to the instructions. Thus, it facilitates
detecting potential cache side channels derived from implicit information flow.}
In contrast, Rule \textsc{Cond.II} takes the least upper bound of two branch
expressions' types.

Statement type rules are standard (see Appendix~\ref{appedix:rules_stmt}), and we
emphasize that \tool\ tracks secrets propagation through both explicit and implicit 
information flows.

\begin{proposition}
\label{prop:typerule-new}
Our type system guarantees security-safety statically: if an expression $e$ is
given the type $\{v: T \mid v: \tau\}$, then the type of its runtime value will
be at least at level $\tau$ on the type hierarchy.
\end{proposition}

\rev{That is, the type system in \tool\ is sound, and it does not make any false
negatives in its analysis; see further discussions and empirical results about
type system correctness in \S~\ref{sec:discussion}.}
%
%

\subsection{Cache Side Channel Detection}
\label{subsec:detection}

\S~\ref{subsec:cache-sc} has illustrated two representative forms of cache side
channels, i.e., SDMA and SDBC. When performing type inference, \tool\ will check
each encountered memory access or conditional jump instruction to see if cache
side channels exist. Specifically, to check if a memory access leads to SDMA, we
right shift the variable holding memory address by $L$ bits, and decide
if the resulting variable is of $\sdd$ type. Following a common
setup~\cite{wang2017cached,wang2019identifying,bao2021abacus}, $L$ equals 6,
standing for 64-byte ($2^6$) cache line size on modern CPUs.

For SDBC, previous research~\cite{brotzman2019casym,bao2021abacus} merely checks
if different secrets induce distinct executing branches. In contrast, \tool\
checks if the conditional expression is of $\sdd$ type, and further assures two
branches are not within identical cache lines. Recall as shown in
\F~\ref{fig:architecture}, we disassemble the crypto software executable and
recover the control flow structure. At this step, we compute the covered cache
units of two branches: a SDBC is confirmed, in case the condition is of $\sdd$
type, and two branches are placed within distinguishable (at least one
non-overlapping) cache lines.

\begin{figure*}[htbp]
\centering
\begin{minipage}[c]{0.3\linewidth}
\centering
\caption{\small \textrm{BN\_num\_bits\_word}.}
\label{list3}
\begin{lstlisting}
804961d: mov eax, ptr [ebp+0x8]
8049620: and eax, 0xffff0000
8049625: test eax, eax
// secret-dependent condition
8049627: je 8049661
8049629: mov eax, ptr [ebp+0x8]
804962c: and eax, 0xff000000
8049631: test eax, eax
// secret-dependent condition
8049633: je 804964b
8049635: mov eax, ptr [ebp+0x8]
8049638: shr eax, 0x18
// secret-dependent mem access
804963b: mov al, ptr [eax+0x8110460]
8049641: and eax, 0xff
8049646: add eax,0x18
8049649: jmp 8049691
\end{lstlisting}
\end{minipage}\hspace{5pt}
\begin{minipage}[c]{0.65\linewidth}
    \centering
    \captionof{table}{\small Type Inference. ``c-line'' stands for cache line.}
    \label{tab:typeexample}
    \footnotesize
    \begin{threeparttable}
  \setlength{\tabcolsep}{2pt}    
\resizebox{1.00\linewidth}{!}{
    \begin{tabular}{|l|l|l|l|}
    \hline
        \multicolumn{1}{|c|}{Involved refinment types}
        &\multicolumn{1}{c|}{Applied rules}
        &\multicolumn{1}{c|}{Control-flow \& cache lines}\\
    \hline
        $eax=\lbrace K \rbrace^{32}:SDD$&
        & \multirow{3}{*}{}\\
    \cline{1-2}
        $eax=\lbrace K \rbrace^{16}\lbrace 0 \rbrace^{16}:SDD,\ r_{0} =\lbrace 1 \rbrace^{16}\lbrace 0 \rbrace^{16}:CST$
        &\textsc{Logic.I}, \textsc{Conj$\&$Disj.I}, \textsc{Const-Conj.I$\&$II}
        &\\
    \cline{1-2}
        \begin{tabular}[c]{@{}l@{}}
            $eax=\lbrace K \rbrace^{16}\lbrace 0 \rbrace^{16}:SDD,\ r_{0}=\lbrace K \rbrace^{16}\lbrace 0 \rbrace^{16}:SDD,$\\
            $zf=\lbrace K \rbrace:SDD$
        \end{tabular}
        &\textsc{Logic.I}, \textsc{Conj$\&$Disj.I}, \textsc{Const-Conj.I}
        &\\
    \cline{1-3}
        $\textcolor{red}{\texttt{je}~\text{condition}~(zf) \longrightarrow \text{secret-dependent}}$
        &
        &
        \multirow{5}{*}{\begin{tabular}[c]{@{}l@{}}
            BC(8049629,8049661,804968c)\\
            $\text{true branch} \longrightarrow \text{c-line 201258}$\\
            $\text{false branch} \longrightarrow \text{c-line 201259 20125a}$
        \end{tabular}}\\
    \cline{1-2}
        $eax=\lbrace K \rbrace^{32}:SDD$
        &&\\
    \cline{1-2}
        $eax=\lbrace K \rbrace^{8}\lbrace 0 \rbrace^{24}:SDD,\ r_{0}=\lbrace 1 \rbrace^{8}\lbrace 0 \rbrace^{24}:CST$
        &\textsc{Logic.I}, \textsc{Conj$\&$Disj.I}, \textsc{Const-Conj.I$\&$II}
        &\\
    \cline{1-2}
        \begin{tabular}[c]{@{}l@{}}
            $eax=\lbrace K \rbrace^{8}\lbrace 0 \rbrace^{24}:SDD,\ r_{0}=\lbrace K \rbrace^{8}\lbrace 0 \rbrace^{24}:SDD,$\\
            $zf=\lbrace K \rbrace:SDD$
        \end{tabular}&
        \textsc{Logic.I}, \textsc{Conj$\&$Disj.I}, \textsc{Const-Conj.I}
        &\\
    \hline
        $\textcolor{red}{\texttt{je}~\text{condition}~(zf) \longrightarrow \text{secret-dependent}}$
        &
        &
        \multirow{3}{*}{\begin{tabular}[c]{@{}l@{}}
            BC(8049635,804964b,804965f)\\
            $\text{true branch}  \longrightarrow \text{c-line 201258}$\\
            $\text{false branch}  \longrightarrow \text{c-line 201259}$
        \end{tabular}}\\
    \cline{1-2}
        $eax=\lbrace K \rbrace^{32}:SDD$&
        &\\
    \cline{1-2}
        $eax=\lbrace 0 \rbrace^{24}\lbrace K \rbrace^{8}:SDD,\ r_{0}=24:CST$
        &\textsc{Extraction}, \textsc{Concat.I}
        &\\
    \hline
        \begin{tabular}[c]{@{}l@{}}
            $eax=\lbrace 0 \rbrace^{24}\lbrace K \rbrace^{8}:SDD,\ r_{0}=135332960:CST,$\\
            $r_{1}=\lbrace 0 \rbrace^{4}\lbrace 1 \rbrace\lbrace 0 \rbrace^{6}\lbrace 1 \rbrace\lbrace 0 \rbrace^{3}\lbrace 1 \rbrace\lbrace 0 \rbrace^{5}\lbrace 1 \rbrace\lbrace 0 \rbrace^{2}\lbrace K \rbrace^{8}:SDD,$\\
            $\textcolor{red}{\text{memory address}~(r_{1}) \longrightarrow \text{secret-dependent}}$
        \end{tabular}
        &\textsc{Arith.I}, \textsc{Concat.I}
        &
        \begin{tabular}[c]{@{}l@{}}
            MA(804963b)\\
            $\text{destination} \longrightarrow \text{c-line 0x201258} \cdots$
        \end{tabular}
        \\
    \hline
        $eax=\lbrace 0 \rbrace^{24}\lbrace K \rbrace^{8}:SDD,\ r_{0}=\lbrace 0 \rbrace^{24}\lbrace 1 \rbrace^{8}:CST$
        &\textsc{Logic.I}, \textsc{Conj$\&$Disj.I}, \textsc{Const-Conj.I$\&$II}
        &\multirow{3}{*}{}\\
    \cline{1-2}
        $eax=\lbrace 0 \rbrace^{24}\lbrace K \rbrace^{8}:SDD,\ r_{0}=24:CST$
        &\textsc{Arith.I}, \textsc{Concat.I}
        &\\
    \cline{1-2}
        &
        &BR(8049649,8049691) \\
    \hline
    \end{tabular}
}
    \begin{tablenotes}
    \item $\dagger\ r_{0}$ and $r_{1}$ represent temporary variables. $\ddagger\ zf$ represents Zero Flag register.
    \end{tablenotes}
    \end{threeparttable}
\end{minipage}
\caption{Type inference over sample assembly code. To ease reading, we use K, I,
    W, and U to term refinement type predicates, corresponding to \sdd, \sid,
    \ukd, and \rad\ types. $\lbrace K \rbrace^{32}$ means bit K repeats 32
    times, while $\lbrace 1 \rbrace^{16}$ means bit 1 repeats 16 times.}
\end{figure*}

\parh{An Illustrative Example.}~\rev{We use an example from the OpenSSL library to
visually demonstrate the type inference and detection of side channels. With
respect to code in \F~\ref{list3}, we present the corresponding (simplified)
type inference procedure launched by \tool\ in \T~\ref{tab:typeexample}. The
first and second columns report the applied type inference rules and the
refinement types of relevant variables. The last column reports the relevant
cache line layout: MA($a$) represents a secret-dependent memory access, and we
also report the accessed cache line. BC($a,b,c$) indicates that for a
conditional control transfer the \texttt{if} branch starts at virtual address
$a$ (ends at address $b$), whereas the \texttt{else} branch starts at $b$ and
ends at $c$. We also report the accessed cache lines in the last column (``c-line'').}

\rev{Before analysis, users mark \texttt{eax} as ``secrets'' (type $\sdd$). With
type inference applied, \tool\ identifies one SDMA and two SDBC (marked in red).
As shown in the last column, for the memory address of the SDMA, \tool\ checks that
the refinement type of highest $32-L$ bits is of $\sdd$
type. As for those two SDBC cases, in addition to checking the branch condition's
type is $\sdd$, \tool\ further checks whether the \texttt{if} and \texttt{else}
branches are located within distinguishable cache lines. \tool\ confirms all
three cases as vulnerable to cache side channels, whose findings are aligned
with~\cite{wang2017cached, wang2019identifying}.}

%% file: 5_implementation.tex
\section{Implementation}
\label{sec:implementation}

\rev{\tool\ is implemented in Scala, and presently performs analysis on crypto
software executables compiled on 32-bit x86 platforms. However, extending \tool\
to other platforms, e.g., 64-bit x86, is not complex. See discussion in
\S~\ref{sec:discussion}. As a common practice for trace-based analysis, we use
Pin~\cite{luk2005pin} to log each covered instruction and its associated
execution context, including all values in CPU registers. These logged contexts
are used to compute the concrete values of pointers in the follow-up static
analysis phase. In other words, our type inference phase employs a practical and
common memory model~\cite{wang2017cached,brumley2011bap}, such that we decide
the addresses stored in a pointer using their concrete values logged on the
trace.}

\rev{We use \textit{objdump} to disassemble executable files of crypto software, and 
recover the control flow graph over the disassembled assembly code. Currently,
when encountering an indirect jump, we conservatively consider that it can jump to
any legitimate control transfer destinations in the disassembled assembly code.
For each conditional jump, we collect the memory address ranges of its \texttt{if/else}
branches from the disassembled code. We build a lookup table over these control
transfer information when checking if executing secret-dependent branches can
visit different cache lines.}

\parh{Usage of \tool.}~\rev{To use \tool, users need to manually identify the secrets
and random factors like blinding in assembly code of crypto software. As noted
in \S~\ref{subsec:assumption}, \tool\ is designed primarily for crypto software
developers, who have detailed knowledge of their own code. Note that the
knowledge of sensitive data in crypto binary code is generally assumed by
previous side channel detectors, as most of them analyze binary
code~\cite{wang2017cached,wang2019identifying,bao2021abacus,weiser2018data}.}

\rev{We clarify that, as existing
works~\cite{bao2021abacus,wang2017cached,wang2019identifying}, flagging secret
(e.g., RSA private key) only requires mundane reverse engineering of crypto
executable and marking memory buffers that store keys. To date, disassemblers
are mature for processing crypto executables. Moreover, to ease the localization
of secrets/random factors in assembly code, we recommend developers to compile
crypto software with debug information attached. We observe that it takes less
than 30 minutes to flag the secrets for each of our evaluated crypto software.
Other than manually localizing secrets, all follow-up analyses are done
automatically by \tool, whose outputs would be localized vulnerable points in
assembly code, as illustrated in \T~\ref{tab:typeexample}. Then, developers will
need to map those leakage assembly instructions to source code for diagnosis and
patching. To ease mapping assembly instructions to source code, it is also
suggested to compile binary code with debug information attached, thereby
encoding source code line number into assembly instructions.}

\rev{In addition, we do not particularly mark certain one-way
functions on the execution trace, e.g., functions applying key blinding over
secrets.
Instead, we
assign refined types (\rad) to random data before the
analysis, and whenever keys are used together with blinding, refined types
for secrets and blinding will naturally fit their corresponding type
inference rules (as defined in \F~\ref{fig:bitrule-new} and
\F~\ref{fig:typerule-new}). Therefore, we should not miss any one-way function
provided that random data has been marked correctly before the analysis.}

%% file: 6_evaluation.tex
\section{Evaluation}
\label{sec:evaluation}

\parh{Evaluation Setup.}~We evaluate \tool\ on production cryptosystems.
Evaluations are conducted in Ubuntu 16.04 with Intel Xeon 3.50GHz CPU, 32GiB
RAM. We collect execution traces of algorithms including RSA, Elgamal, and
(EC)DSA from OpenSSL and Libgcrypt (see \T~\ref{tab:cryptosystems}). $*$
represents using random factor on plaintext/ciphertext and $\star$ indicates
using random factor on secrets. Besides, we evaluate the effectiveness of \tool\
on a constant-time dataset offered in Binsec/Rel~\cite{daniel2020binsec}. This
will validate the correctness of our methodology to a reasonable extent.

The RSA/Elgamal algorithms from both libraries leverage the built-in secret
generation function for generating 2048-bit secrets. The ECDSA algorithm adopts
OpenSSL \textit{sect571r1} curve. We initiate the plaintext or the message to be
signed as ``hello world''. We use Intel Pin to log the execution traces when
executing the crypto software for standard decryption/signature procedures,
including the majority of asymmetric encryption functions such as modular
exponentiation in RSA/ElGamal and point multiplication in the signature
procedure of ECDSA.

\begin{table}[htbp]
\centering
\caption{Cryptosystems analyzed by \tool.}
\label{tab:cryptosystems}
\footnotesize
  \resizebox{0.80\linewidth}{!}{
\begin{tabular}{|c|c|c|}
\hline
    \textbf{Algorithms}
    &\textbf{Implementations}
    &\textbf{Versions} \\
\hline
    \multirow{2}{*}{RSA}
    &OpenSSL
    &\begin{tabular}[c]{@{}c@{}}$1.0.2f^{*}$, $1.1.0g^{*}$, $1.1.0h^{*}$\\ $1.1.1n^{*}$, $3.0.2^{*}$\end{tabular}\\
    \cline{2-3}
    &Libgcrypt
    &$1.6.1^{*}$, $1.7.3^{*}$, $1.9.4^{*\star}$ \\
\hline
    ElGamal
    &Libgcrypt
    &1.6.1, $1.7.3^{*}$, $1.9.4^{*\star}$ \\
\hline
    (EC)DSA
    &OpenSSL
    &\begin{tabular}[c]{@{}c@{}}1.0.1e, 1.1.0g, $1.1.0i^{\star}$\\ $1.1.1n^{\star}$, $3.0.2^{\star}$\end{tabular} \\
\hline
\end{tabular}
  }
\end{table}

\subsection{Results Overview}
\label{subsec:results}

\begin{table*}[htbp]
\centering
\caption{Identified Information Leakage Sites/Units by \tool. We compare the
results with recent works, including CacheD~\cite{wang2017cached},
CacheS~\cite{wang2019identifying} and DATA~\cite{weiser2018data,
weiser2020big}.}
\label{tab:resultsoverview}
\begin{threeparttable}
\scriptsize
\begin{tabular}{c|c|c|c|c|c|c}
\hline
    \multirow{2}{*}{\textbf{Algorithms}}
    &\multirow{2}{*}{\textbf{Implementations}}
    &\multirow{2}{*}{\textbf{\begin{tabular}[c]{@{}c@{}}Information Leakage\\ Sites (known/unknown)\end{tabular}}}
    &\multirow{2}{*}{\textbf{\begin{tabular}[c]{@{}c@{}}Information Leakage\\ Units (known/unknown)\end{tabular}}}
    &\textbf{CacheD reported \cite{wang2017cached}}
    &\textbf{CacheS reported \cite{wang2019identifying}}
    &\textbf{DATA reported \cite{weiser2018data, weiser2020big}}\\
    \cline{5-7}
    & & &
    &\textbf{Leakage Sites/Units}$\dagger$
    &\textbf{Leakage Sites/Units}$\dagger$
    &\textbf{Leakage Units}$\ddagger$\\
\hline
    \textbf{RSA}
    & OpenSSL 1.0.2f
    & 30/0 & 6/0 & 2/2 & 6/3 & 4\\
    \textbf{RSA}
    & OpenSSL 1.1.0g
    & 30/4 & 8/1 & - & - & 5\\
    \textbf{RSA}
    & OpenSSL 1.1.0h
    & 22/0 & 5/0 & - & - & 5\\
    \textbf{RSA}
    & OpenSSL 1.1.1n
    & 9/0 &5/0 & - & - & 3\\
    \textbf{RSA}
    & OpenSSL 3.0.2
    & 9/4 & 4/2 & - & - & 2\\
    \textbf{RSA}
    & Libgcrypt 1.6.1
    & 31/4 & 9/1 & 22/5 & 40/11 & -\\
    \textbf{RSA}
    & Libgcrypt 1.7.3
    & 24/4 & 8/1 & 0/0 & 0/0 & -\\
    \textbf{RSA}
    & Libgcrypt 1.9.4
    & 4/5 & 2/3 & - & - & -\\
\hline
    \textbf{ElGamal}    
    & Libgcrypt 1.6.1
    & 31/4 & 9/1 & 22/5 & 40/11 & -\\
    \textbf{ElGamal}    
    & Libgcrypt 1.7.3
    & 24/4 & 8/1 & 0/0 & 0/0 & -\\
    \textbf{ElGamal}    
    & Libgcrypt 1.9.4
    & 3/0 & 1/0 & - & - & -\\
\hline
    \textbf{ECDSA}
    & OpenSSL 1.0.1e
    & 98/0 & 9/0 & - & - & 9\\
    \textbf{ECDSA}
    & OpenSSL 1.1.0g
    & 49/0 & 6/0 & - & - & 6\\
    \textbf{ECDSA}
    & OpenSSL 1.1.0i
    & 13/0 & 3/0 & - & - & 3\\
    \textbf{ECDSA}
    & OpenSSL 1.1.1n
    & 14/0 & 2/0 & - & - & 2\\
    \textbf{ECDSA}
    & OpenSSL 3.0.2
    & 14/0 & 2/0 & - & - & 3\\
\hline
    \textbf{DSA}$\natural$
    & OpenSSL 1.1.0i
    & 0/4 & 0/1 & - & - & -\\
    \textbf{DSA(swapped)}$\natural$
    & OpenSSL 1.1.0i
    & 9/4 & 4/1 & - & - & -\\
    \textbf{DSA}
    & OpenSSL 1.1.1n
    & 13/4 & 3/1 & - & - & 3\\
    \textbf{DSA}
    & OpenSSL 3.0.2
    & 13/4 & 3/1 & - & - & 3\\
\hline
    \textbf{total}
    &
    & 440/45 & 97/14 & 46/12 & 86/25 & 48\\
\hline
\end{tabular}
\begin{tablenotes}
\item $\dagger$ The RSA and Elgamal from Libgcrypt library are counted together
in CacheD~\cite{wang2017cached} and CacheS~\cite{wang2019identifying}.

\item $\ddagger$ We collect all leaky functions reported in
DATA~\cite{weiser2018data, weiser2020big} and locate whether these leaky
functions appear in the corresponding OpenSSL version.

\item $\natural$ DSA (OpenSSL-1.1.0i) and its swapped patch are only evaluated
for the key blinding part.
\end{tablenotes}
\end{threeparttable}
\end{table*}

\begin{figure*}[htbp]
    \centering
    \begin{minipage}[c]{0.6\linewidth}
        \centering
        \captionof{table}{Performance comparison with CacheD/CacheS. We also
        list the analysis of eight RSA implementations for scalability
        assessment.}
        \label{tab:performanceoverview}
        \scriptsize
        \begin{tabular}{c|c|c|c|c|c}
        \hline
            \multirow{2}{*}{\textbf{Crypto setup}}
            &\multirow{2}{*}{\textbf{\begin{tabular}[c]{@{}c@{}}Instructions\\ on the Traces\end{tabular}}}
            &\multirow{2}{*}{\textbf{\begin{tabular}[c]{@{}c@{}}Processing Time\\ (CPU Seconds)\end{tabular}}}
            &\multirow{2}{*}{\textbf{\begin{tabular}[c]{@{}c@{}}Time of\\ Per $10^{4}$ Lines\end{tabular}}}
            &\textbf{CacheD}
            &\textbf{CacheS}\\
            \cline{5-6}
            & & &
            &\textbf{Per $10^{4}$ Lines}
            &\textbf{Per $10^{4}$ Lines}\\
        \hline
            \textbf{\begin{tabular}[c]{@{}c@{}}RSA \& Elgamal\\ OpenSSL-1.0.2f\end{tabular}}
            & 1,620,404
            & 35.58
            & 0.22
            & 3.49
            & 21.16\\
        \hline
            \textbf{\begin{tabular}[c]{@{}c@{}}RSA \& Elgamal\\ Libgcrypt-1.6.1\end{tabular}}
            & 1,379,652
            & 36.00
            & 0.26
            & 4.93
            & 45.36\\
        \hline
            \textbf{\begin{tabular}[c]{@{}c@{}}RSA \& Elgamal\\ Libgcrypt-1.7.3\end{tabular}}
            & 1,411,081
            & 48.40
            & 0.34
            & 3.92
            & 54.57\\
        \hline
            \textbf{total (first three rows)}
            & 4,411,137
            & 119.98
            & 0.27
            & 4.42
            & 35.41\\
        \hline 
            \textbf{RSA-OpenSSL 1.0.2f}
            & 1,620,404
            & 35.58
            & 0.22
            & -
            & -\\
            \textbf{RSA-OpenSSL 1.1.0g}
            & 822,151
            & 18.58
            & 0.22
            & -
            & -\\
            \textbf{RSA-OpenSSL 1.1.0h}
            & 28,874
            & 4.88
            & 1.69
            & -
            & -\\
            \textbf{RSA-OpenSSL 1.1.1n}
            & 1,763,970
            & 39.29
            & 0.22
            & -
            & -\\
            \textbf{RSA-OpenSSL 3.0.2}
            & 1,711,746
            & 36.57
            & 0.21
            & -
            & -\\
            \textbf{RSA-Libgcrypt 1.6.1}
            & 806,410
            & 22.63
            & 0.28
            & -
            & -\\
            \textbf{RSA-Libgcrypt 1.7.3}
            & 837,215
            & 23.23
            & 0.27
            & -
            & -\\
            \textbf{RSA-Libgcrypt 1.9.4}
            & 114,733
            & 11.25
            & 0.98
            & -
            & -\\
        \hline
        \end{tabular}
    \end{minipage}\hspace{10pt}
    \begin{minipage}[c]{0.35\linewidth}
        \centering
        \includegraphics[width=0.95\textwidth]{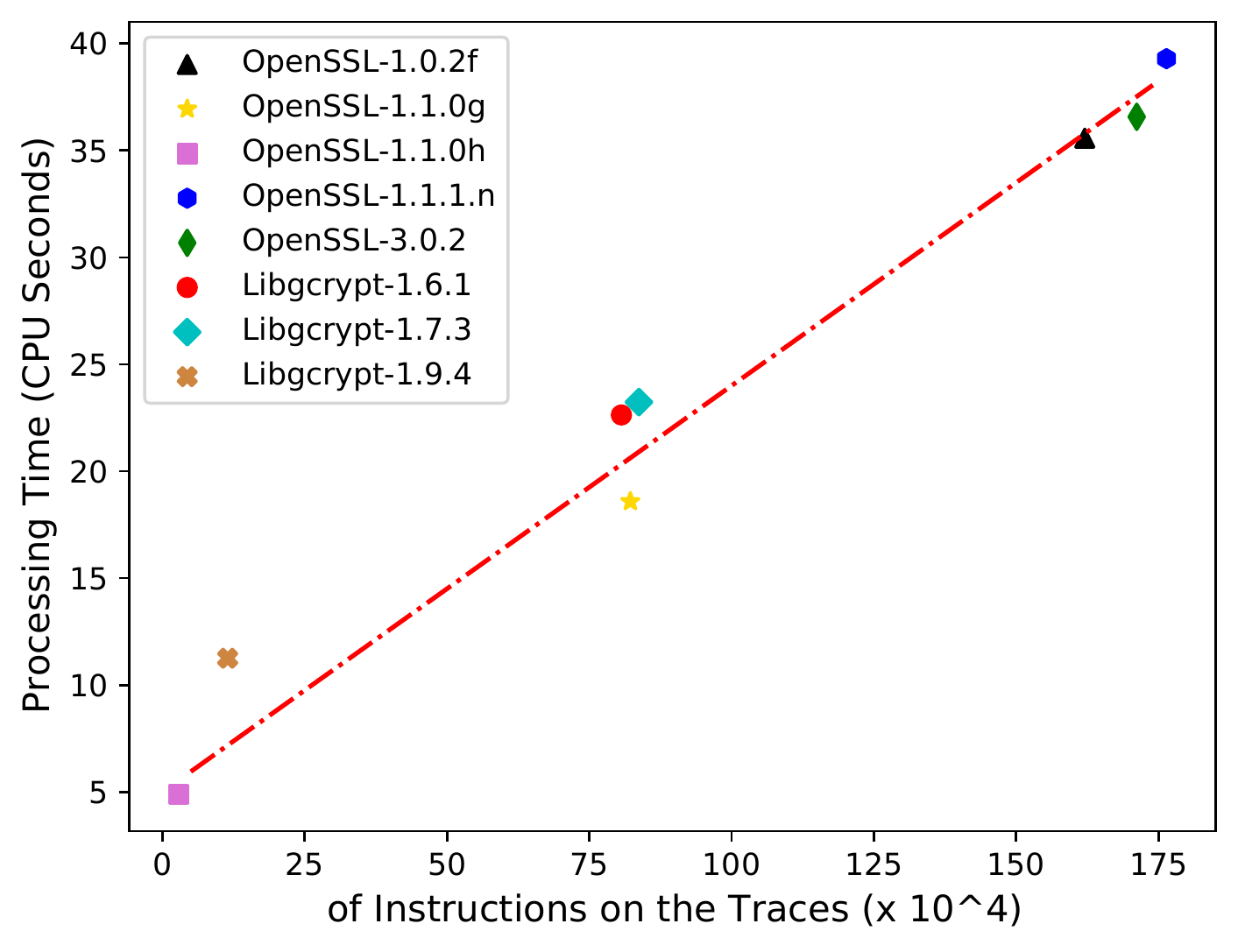}
        \caption{Trace lengths/processing time towards the analysis of RSA implementations.}
        \label{fig:rsaperformance}
    \end{minipage}
    \vspace{-5pt}
\end{figure*}

\parh{Vulnerability Detection.}~We present the positives reported by \tool\ in
\T~\ref{tab:resultsoverview}. \rev{We report that \tool\ confirms all cache side
channel vulnerabilities that have been found by CacheD/CacheS. Moreover, it
identifies new defects that were neglected in previous analyses of the same
crypto software.} \tool\ detects precisely 485 information leakage sites,
including 440 known sites and 45 newly found sites. To better characterize
findings, we adhere to CacheD/CacheS to group adjacent leakage sites (assembly
instructions) into a unit and eliminate duplicated units. This way, 97 known
units are confirmed and 14 unknown units are discovered.
\cite{weiser2018data,weiser2020big} only report leakage units, which are
compared here. We elaborate on the findings of \tool\ in the following two
subsections. 

Also, for the constant-time dataset offered by~\cite{daniel2020binsec}, \tool\
has \textit{no} positive findings, meaning that \tool\ (over this dataset) does not
produce false positives or false negatives. We notice that constant-time
computations in this dataset (e.g., comparison and conditional selection)
extensively use bitwise operations. Since \tool\ performs bit-level type
inference, \tool\ manifests high accuracy without treating safe bitwise
operations as vulnerable. Note that constant-time operations provided in this
dataset are frequently used in modern crypto libraries; thus, experiments on
this dataset verify the correctness of \tool\ to a reasonable extent.

\parh{Analysis Against Randomization.}~\tool\ is evaluated against blinding
over plaintext/ciphertext and keys. \tool\ confirms that the secret leakage
exists in OpenSSL-1.0.2f and Libgcrypt-1.6.1/1.7.3, notwithstanding the
introduction of plaintext/ciphertext blinding. Note that secrets are still
exposed to side channels without blinding in these cases. In contrast, key
blinding mitigates most leakage sites. For instance, evaluations of RSA/ElGamal
in Libgcrypt-1.9.4 reveal that secrets are now labeled as random data (with type
\ura) by \tool. However, this protection is at the cost of introducing extra
(potentially vulnerable) procedures to perform blinding. \tool\ discovers five
new leakage sites in RSA/Libgcrypt-1.9.4. These leakage units cover both the
private key $d$ and the prime $p$ (recall in RSA, $d$ and $p$ are secrets).
Therefore, we show that though key blinding obscures secrets, it introduces new
leakage sites due to extra calculations. In sum, by considering random factors
with specific refined types, \tool\ can analyze side channel mitigation
techniques implemented in modern crypto software.

\parh{Performance Evaluation.}~We compare \tool\ with CacheD and CacheS by using 
the same crypto implementations, and report the comparison results in
\T~\ref{tab:performanceoverview} (first five rows). 
For crypto libraries evaluated by CacheD/CacheS (with a total of 4.4M
instructions), \tool\ finishes the analysis with around 120 CPU seconds,
\rev{and exhibits promising speed across all evaluation settings with no
timeout cases.} To compare with CacheD/CacheS, we use the processing time per 10
thousand lines as an indicator. \tool\ handles per 10 thousand lines in 0.27
seconds on average, while CacheD and CacheS require 4.42 CUP and 35.41 CPU
seconds, respectively. We also report performance statistics of other RSA
evaluation settings in the next rows of \T~\ref{tab:performanceoverview}. Their
trace lengths range between thousands and millions. \F~\ref{fig:rsaperformance}
illustrates the approximately linear correlations between trace length and time.
Considering the complexity of analyzing real-world cryptosystems, \tool\
displays a highly promising performance and scalability.

The performance comparison results (\T~\ref{tab:performanceoverview}) demonstrate
the superiority of type inference as opposed to existing works (e.g.,~\cite{wang2017cached,wang2019identifying,bao2021abacus,brotzman2019casym}) that use the constraint solver to decide the satisfiability of side channel constraints.
Holistically, those works suffer from the accumulation of complex
constraints when performing symbolic execution along the trace. 
In contrast, type inference ensures each deduction step has a straightforward
result without huge search space. 
Overall, without using constraint solving, \tool\ maintains a comparable analysis capability as those of CacheD/CacheS.
As noted in \S~\ref{subsec:detection}, by using bit-level secret tracking
($\sdd$), deciding if secret-dependent memory access leads to cache side
channels is recast to essentially recognize $\sdd$ in refined types. This pattern match
operation is very efficient without undermining soundness.

\subsection{Discussion of Known Vulnerabilities}
\label{knownfinds}

\tool\ confirms all vulnerabilities reported by CacheD/CacheS in the RSA/Elgamal
implementations from Libgcrypt-1.6.1, which adopts pre-computation tables for
the sliding-window exponentiation (see \F~\ref{fig:vulknownlibg} of
Appendix~\ref{appdix:known}). Although Libgcrypt-1.7.3 employs a direct
computation scheme rather than using pre-computation tables, \tool\ still
finds 24 leakage sites that leak the secret length, which also exist in
Libgcrypt-1.6.1. However, no leaks are reported in CacheD/CacheS about
Libgcrypt-1.7.3. In the CacheS paper, they admit these leak points are false
negatives of their tool. As Libgcrypt-1.9.4 adopts a new algorithm (i.e.,
left-to-right exponentiation), \tool\ reports a known secret length leakage in
function \textrm{\_gcry\_mpih\_add\_n}, whereas prior leak operations are
discontinued.

Concerning OpenSSL, \tool\ first confirms the existence of CVE-2018-0737, where
RSA private key is leaked during key generation, in functions \textrm{BN\_gcd}
and \textrm{BN\_mod\_inverse} from OpenSSL-1.1.0g/1.1.0h. In analyzing modular
inverse, \tool\ detects a new vulnerability in the function \textrm{BN\_rshift1}
that discloses the length of the secret (see \S~\ref{newfound}). A recently
found vulnerability comes from function \textrm{BN\_num\_bits\_word}, reported
in CacheD/CacheS (See \F~\ref{fig:vulknownrsa} of
Appendix~\ref{appdix:known}). \tool\ performs the type deduction process in
\T~\ref{tab:typeexample}. The issue exists in OpenSSL-1.0.2f/1.1.0g/1.1.0h and
has been fixed~\cite{openssl@972c87d},
hence disappears in the latest OpenSSL versions (OpenSSL-1.1.1n/3.0.2). \tool\
confirms a secret length leakage in function
\textrm{BN\_window\_bits\_for\_ctime\_exponent\_size} in all analyzed OpenSSL
versions, shown in Listing~\ref{listrsa}. The issue is also reported in CacheS,
but is not fixed in the latest OpenSSL. \tool\ also detects a vulnerability
reported in DATA, where constant-time flags of RSA secret
primes $p$ and $q$ are not propagated to the temporary copies inside the
function \textrm{BN\_MONT\_CTX\_set} during the Montgomery initialization for
modular inverse. This issue exists in OpenSSL-1.0.2f, but the other four OpenSSL
libraries resolve it.


{
\begin{lstlisting}[language=C, frame=single, basicstyle=\scriptsize, caption=Window size of modular exponentiation., label=listrsa]
1 BN_window_bits_for_ctime_exponent_size(b) \
2       ((b) > 937 ? 6 : \
3       (b) > 306 ? 5 : \
4       (b) >  89 ? 4 : \
5       (b) >  22 ? 3 : 1)
\end{lstlisting}
}

When evaluating the (EC)DSA implementations, we mark the \texttt{nonce} used in
Montgomery ladder as a secret. This is because the leaky nonce can result in the
Hidden Number Problem (HNP)~\cite{boneh1996hardness, boneh1997rounding}, where
collecting enough leaky nonce contributes to the recovery of private keys
through constructing lattice~\cite{benger2014ooh, nguyen2002insecurity,
nguyen2003insecurity}. \tool\ confirms a direct leakage of the nonce in the
Montgomery ladder implementation from OpenSSL-1.0.1e. This vulnerability was 
reported in~\cite{yarom2014recovering},
and this flaw (CVE-2014-0076) has been fixed by the developers and implemented
in a non-branch commit~\cite{openssl@4b7a4ba, openssl@2198be3}.
Recently, Ryan~\cite{ryan2019return} reports a vulnerability located in modular
reduction of (EC)DSA implementations in OpenSSL that uses an early abort
condition to estimate the range of private keys. \tool\ confirms this
vulnerability comes from function \textrm{BN\_ucmp} and \textrm{BN\_usub} inside
function \textrm{BN\_mod\_add\_quick} (see \F~\ref{fig:vulknownecdsa} of
Appendix~\ref{appdix:known}).

\tool\ is also evaluated on analyzing the lifetime of a nonce, including the
generation, scalar multiplication, modular inversion, and main signing process.
The leakage sites identified by \tool\ fully cover the findings reported
in~\cite{weiser2020big}. For example, by distinguishing whether an extra limb is
used to expand the representation of nonce in \textrm{BN\_add}, \tool\ confirms
the padding resize vulnerabilities about the nonce reported in CVE-2018-0734 for
DSA and CVE-2018-0735 for ECDSA, as shown in Listing~\ref{listecdsa}. The
vulnerability states that the result buffer resizes one more limb to hold the
result. By distinguishing the resize operations, attackers can learn the range
information of the nonce. Other known leakage sites of the nonce (e.g., skipping
leading zero limbs through \textrm{bn\_correct\_top}, performing an early stop
in \textrm{BN\_cmp}, and conditional branches in \textrm{BN\_mul}) are
identified by \tool; they still exist in the latest versions. Individually,
\tool\ reports non-constant-time vulnerabilities in OpenSSL-1.0.1e when
performing ECDSA nonce modular inverse. This is because the constant-time flag
was not set to the nonce. OpenSSL-1.1.0g/1.1.0i, on the other hand, implement
Fermat's little theorem via constant-time modular exponentiation. Benefit to the
cache layout checking, \tool\ finds four new leakage sites that reveal the secret
key size through a series of \texttt{else/if} branches in DSA from
OpenSSL-1.1.0i/1.1.1n/3.0.2 (see \S~\ref{newfound}). Contrary to our
expectations, \tool\ does not mark cases in the \texttt{switch} statement of \textrm{BN\_copy}
as vulnerable. Through rechecking the source code and its disassembly code,
we confirm that \tool\ performs a correct inference because the trace on the
cache cannot be distinguished (see \S~\ref{avoidfp}).


{
\begin{lstlisting}[language=C, frame=single, basicstyle=\scriptsize, caption=Bignumber resize., label=listecdsa]
1 if (!BN_add(r, k, order)
2  || !BN_add(X, r, order)
3  || !BN_copy(k, BN_num_bits(r)>order_bits ? r:X))
4  goto err;
\end{lstlisting}
}

\subsection{Unknown Vulnerabilities}
\label{newfound}

\tool\ finds new vulnerable program points in Libgcrypt-1.6.1/1.7.3 that have
been analyzed by existing tools. It finds that the size of secret exponentiation
is leaked through the \texttt{if/else} statements at the beginning of function
\textrm{\_gcry\_mpi\_powm}, as shown in Listing~\ref{listlibg}. The
sliding-window size \texttt{W} is determined by the size of secret exponent
\texttt{esize}. Different execution traces of the \texttt{if/else} statements can be
differentiated because it occupies multiple cache lines. However, we admit that
the \texttt{if/else} statements are a moderate leakage because only line 1 and line 5 can
be distinguished directly. \tool\ cannot distinguish execution between line 2
and line 4.

{
\begin{lstlisting}[language=C, frame=single, basicstyle=\scriptsize, caption=Window size selection., label=listlibg]
1 if (esize * BITS_PER_MPI_LIMB > 512) W = 5;
2 else if (esize * BITS_PER_MPI_LIMB > 256) W = 4;
3 else if (esize * BITS_PER_MPI_LIMB > 128) W = 3;
4 else if (esize * BITS_PER_MPI_LIMB > 64) W = 2;
5 else  W = 1;
\end{lstlisting}
}

We find a new vulnerability in the OpenSSL function \textrm{BN\_rshift1} which
performs GCD using the Euclid algorithm. \F~\ref{fig:vulunknown1} in
Appendix~\ref{appdix:unknown} presents the source code from version 1.1.0g. We
first demonstrate how this function leaks the length of the one-shifted-right
operand. Function \textrm{BN\_rshift1} performs shifting to the right one-bit
for each element of the \textrm{BN\_ULONG} structure. The length of the source
operand (i.e., \texttt{a->top}) is used as the while loop's condition. \tool\
confirms it as a secret-dependent branch, where the judgment of the while loops
and part instructions inside the while loops (lines 11--13) are stored in one
cache line and the subsequent instructions until the end of function
\textrm{BN\_rshift1} (lines 14--17) are stored in another cache line.
Therefore, the trace of the while loops can be distinguished. By probing the
while loop condition, the value of \texttt{a->top} is inferred as one increment to the
number of while loops.

\cite{weiser2018single} proposes a page-level attack to recover RSA primes $p$
and $q$ when performing prime testing using \textrm{BN\_gcd}. \tool\ confirms
the vulnerability in which four different branches are identified because
\textrm{BN\_rshift1} of each branch is at different cache lines. Meanwhile, we
argue the length information of the source operand leaked by
\textrm{BN\_rshift1} accelerates the recovery in~\cite{weiser2018single}. For
example, between two adjacent loop operations
($a_{i+1}=a_{i}/2,\ a_{i+1}=(a_{i}-b_{i})/2$ or $a_{i+1}=(a_{i}-b_{i})/2,\
a_{i+1}=a_{i}/2$), one decrement in the latter $a_{i+1}$'s length indicates that
the topmost bit of the former $a_{i+1}$ is one. This deduction helps to reduce
the range of intermediate results for each Euclid loop. In addition to
\textrm{BN\_rshift1}, \tool\ finds similar leakage in \textrm{BN\_lshift1} from
OpenSSL-3.0.2.

\tool\ also finds another vulnerability in the OpenSSL-1.1.0i implementation of
DSA (\F~\ref{fig:vulunknown2} in Appendix~\ref{appdix:unknown}). 
The
key blinding mechanism of DSA first multiplies the random factor \texttt{blind} and the DSA secret
key \texttt{dsa->priv\_key} by calling the function \textrm{BN\_mul}, which calls the function \textrm{bn\_mul\_normal} to perform a classic multiplication if the length of both
operands is less than \textrm{BN\_MULL\_SIZE\_NORMAL}. 
%
\F~\ref{fig:vulunknown2} presents a for-loop, where four elements of the secret 
key as a group are multiplied by \texttt{blind}.
Then, four branches, where \texttt{nb} is the length of the secret key, 
control whether to end the loop.
When \texttt{nb} equals zero, the multiplication is complete and the function 
\textrm{bn\_mul\_normal} returns. \tool\ confirms that the secret-dependent branches 
leak the length of the secret key. By probing different if-conditions present in 
distinct cache lines, the value of the secret length can be recovered. Such vulnerable
operations are found in the latest OpenSSL-1.1.1n/3.0.2.

\subsection{Discussion about Blinding}
\label{blinding}

As stated in \S~\ref{subsec:results}, \tool\ shows that the plaintext/ciphertext
blinding cannot eliminate cache side channels, given that secrets themselves are
still exposed (e.g., secret-dependent memory accesses and branches in modular
exponentiation from Libgcrypt-1.6.1). However, key blinding impedes nearly all
leakage. For example, \tool\ reports no vulnerability in the modular
exponentiation from Libgcrypt-1.9.4. By inspecting the type inference outputs,
we find that the secret exponent is marked as a random number ($\ura$) through a
series of blinding operations before conducting modular exponentiation. However,
\tool\ finds new leakage sites in the blinding process. Considering key blinding
in RSA/Libgcrypt-1.9.4, which uses $d\_blind = (d\ mod\ (p-1)) + (p-1) * r$ to
mask the secret exponent $d$ before performing modular exponentiation. Here, $p$
represents one RSA prime number and $r$ is the random factor. \tool\ newly
discovers five leakage sites in the subtraction and division operations. They
leak the length of the prime number $p$ and secret exponent $d$. For instance,
the function \textrm{\_gcry\_mpi\_sub\_ui} is invoked to perform $p-1$ on $p$.
It leaks the length of $p$ whenever the resize operation is performed on the
result operand, as well as at other length-related branches.

Apart from the key blinding in Libgcrypt-1.9.4, \tool\ also explores the effect
of different key blinding positions on mitigating cache side channels. For
instance, DSA implementation from OpenSSL-1.1.0i applies key blinding \textit{b}
to avoid leaking the private key \textit{x} as follows:

{
\begin{minipage}{0.2\textwidth}
\begin{equation}
s = (bm + bxr)\ mod\ q\label{eq:swap0}
\end{equation}
\end{minipage}
\begin{minipage}{0.2\textwidth}
\begin{align}
s = s \cdot k^{-1}\ mod\ q\label{eq:swap1}\\
s = s \cdot b^{-1}\ mod\ q\label{eq:swap2}
\end{align}
\end{minipage}
}

\noindent where statements~\ref{eq:swap0},~\ref{eq:swap1}, and~\ref{eq:swap2}
are executed sequentially. Swapping statements~\ref{eq:swap1} and~\ref{eq:swap2}
results in different key blinding use, which is applied in a LibreSSL
patch~\cite{libressl@2cd28f9}.
\tool\ compares the original patch with the swapped one (we manually swap
statements~\ref{eq:swap1} and~\ref{eq:swap2} in OpenSSL-1.1.0i DSA). We find
nine additional leakage sites related to the length of the inverse nonce
\texttt{kinv} in the swapped patch (see \T~\ref{tab:resultsoverview}), although
the statement~\ref{eq:swap1} also leaks the inverse nonce length in the original
patch. We argue when executing statement~\ref{eq:swap2} first, $s$ does not
possess the property of randomization anymore due to $b(m + xr)b^{-1}\ mod\ q
\equiv (m + xr)\ mod\ q$. Hence, the nonce inverse \texttt{kinv} is exposed to
the attacker. The swapped practice is fix in a LibreSSL patch~\cite{libressl@1f6b35b}.

\subsection{Reducing False Positives}
\label{avoidfp}

\begin{figure}[htbp]
    \begin{minipage}[c]{0.4\linewidth}
        \begin{subfigure}[b]{0.9\linewidth}
        \centering
        \footnotesize
        \begin{tabular}{ll}
    		&$1\ BIGNUM\ \ast BN\_copy$\\
    		&$2\ \quad (BIGNUM \ast a,$\\
    		&$3\ \quad BIGNUM \ast \textcolor{red}{b}) \lbrace$\\
    		&$4\ \quad \cdots$\\
            &$5\ \quad \scriptsize /*\ assign\ values$\\
            &$6\ \quad \quad \scriptsize in\ groups\ of\ 4\ */$\\
            &$7\ \quad \cdots$\\
            &$8\ \quad switch\ (\textcolor{red}{b->top} \& 3) \lbrace$\\
    		&$9\ \quad case\ 3:\ A[2] = B[2];$\\
    		&$10\ \quad case\ 2:\ A[1] = B[1];$\\
    		&$11\ \quad case\ 1:\ A[0] = B[0];$\\
    		&$12\ \quad case\ 0:; \rbrace$\\
    		&$13\ \quad \cdots \rbrace$\\
    	\end{tabular}
	    \caption{BN\_copy function.}
	    \end{subfigure}
    \end{minipage}\hspace{5pt}
    \begin{minipage}[c]{0.55\linewidth}
        \begin{subfigure}[b]{0.9\linewidth}
            \centering
            \includegraphics[width=0.9\linewidth]{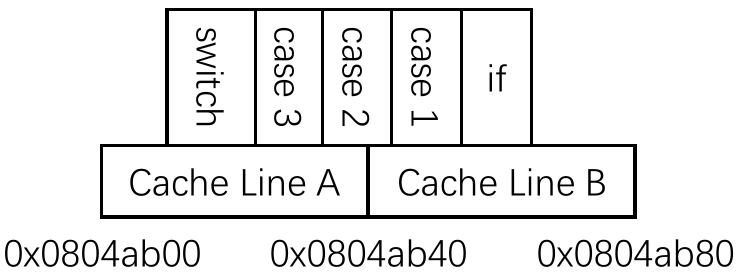}
            \caption{of from OpenSSL-1.1.0g}
        \end{subfigure}\vspace{5pt}
        \begin{subfigure}[b]{0.9\linewidth}
            \centering
            \includegraphics[width=0.9\linewidth]{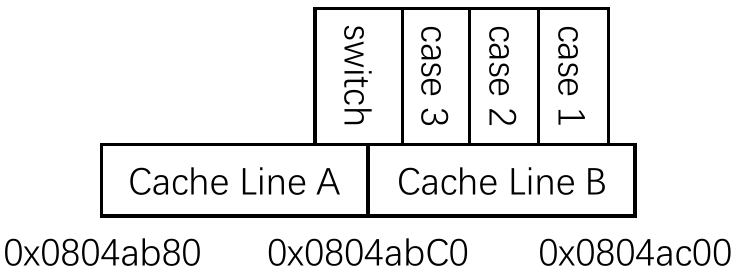}
            \caption{of from OpenSSL-1.1.0h}
        \end{subfigure}
    \end{minipage}
    \caption{BN\_copy from the OpenSSL Library.}
    \vspace{-10pt}
    \label{fig:fps}
\end{figure}

We explain how \tool\ reduces false positives by using cache layouts rather than
cache states to detect side channels. Considering \F~\ref{fig:fps}, function \textrm{BN\_copy} is used by RSA and (EC)DSA. Take (EC)DSA as an
example, whose secret nonce is copied from \texttt{b} to \texttt{a} via
\textrm{BN\_copy}. 
In particular, 
a \texttt{switch} statement at line 8 helps skipping the copy of leading zero in
\texttt{b}. By manually reviewing this function, we would anticipate that 
certain information about the nonce is leaked by discriminating executed switch
cases. However, \tool\ deems this case as safe. 

We analyze the result released by \tool\ from the perspectives of both
$\textsc{FLUSH-RELOAD}$ and $\textsc{PRIME-PROBE}$ attacks. We depict the cache
layouts of \textrm{BN\_copy} from OpenSSL-1.1.0g and OpenSSL-1.1.0h in
\F~\ref{fig:fps}(b) and \F~\ref{fig:fps}(c). In these two libraries, the \texttt{switch}
statement occupies two separate cache lines. Thus, the first cache line
must be visited. Meanwhile, instructions after the statement are loaded
into the second cache line and are also visited; in an extreme case, the whole
\texttt{switch} statement is loaded into one cache line. In sum, different switch cases
are not distinguishable (e.g., for the $\textsc{FLUSH-RELOAD}$ attack). We
further consider whether a $\textsc{PRIME-PROBE}$ attack can distinguish the
difference in cache layouts. First, the base addresses are loaded into the cache
regardless of whether they correspond to the source array (\texttt{A[]}) or the
destination array (\texttt{B[]}). Second, the largest offset for the element
among the last group (both destination and source) is 8 bytes. In that sense,
the address of any element is mapped to the same cache line (address $\gg$ 6 for
64-byte cache lines). Therefore, $\textsc{PRIME-PROBE}$ cannot collect a
distinguishable observation and fails to extract secrets. However, CacheD/CacheS
simply treats \textrm{BN\_copy} as vulnerable, given that a secret-dependent
branch condition (line 8) is (inaccurately) treated as ``vulnerable'' in the
view of their cache state-based vulnerability pattern. However, it is indeed
a false positive.

\parh{Robustness of Using Cache Layouts.}~\rev{The above experiments are
conducted using OpenSSL's default compilation setting. The \texttt{switch}
statement may be vulnerable, when the code chunk of each switch case occupies
distinct cache lines. Overall, we anticipate that different optimization
settings could result in placing instructions into different cache lines. To
benchmark the robustness of using cache layouts instead of using cache
state-based threat models, we measure how compiler optimizations may influence
the results of \tool, whose results are given in \T~\ref{tab:compiler}. At this
step, we only measure side channels due to SDBC, because we use the cache layout
model to check SDBC. Also, given that we need to manually confirm and compare
each finding across different optimizations, we only select a crypto library
when its SDBC-related source code has visible changes across different versions.
For instance, while we evaluate Libgcrypt 1.6.1, 1.7.3, and 1.9.4 in
\T~\ref{tab:resultsoverview}, we only evaluate versions 1.7.3 and 1.9.4, since
version 1.6.1 appears to be identical with 1.7.3 in terms of those SDBC cases
flagged by \tool.}

\begin{table}[htbp]
\centering
\caption{Branch vulnerabilities identified by \tool\ under gcc \texttt{-O0},
\texttt{-O2}, and \texttt{-O3} optimization settings.}
\label{tab:compiler}
\resizebox{0.70\linewidth}{!}{
    \begin{tabular}{c|c|c|c}
    \hline
        \multirow{2}{*}{\textbf{Crypto setup}}
        &\multicolumn{3}{c}{\textbf{gcc-5.4}}\\\cline{2-4}
        &\multicolumn{1}{c|}{\textbf{-O0}}
        &\multicolumn{1}{c|}{\textbf{-O2}}
        &\multicolumn{1}{c}{\textbf{-O3}}
        \\
    \hline
        \textbf{RSA-OpenSSL 1.1.0g}
        & 27/9 & 24/9 & 24/9 \\
        \textbf{RSA-OpenSSL 1.1.0h}
        & 20/5 & 18/5 & 18/5 \\
        \textbf{RSA/Elgamal-Libgcrypt 1.7.3}
        & 17/7 & 14/7 & 14/7 \\
        \textbf{RSA/Elgamal-Libgcrypt 1.9.4}
        & 6/4 & 6/4 & 6/4 \\
    \hline
        \textbf{ECDSA-OpenSSL 1.1.0g}
        & 38/6 & 22/6 & 19/6 \\
        \textbf{ECDSA-OpenSSL 1.1.0i}
        & 10/3 & 7/3 & 7/3 \\
        \textbf{ECDSA-OpenSSL 3.0.2}
        & 9/2 & 9/2 & 9/2 \\
    \hline
        \textbf{DSA-OpenSSL 1.1.0i}
        & 4/1 & 3/1 & 3/1 \\
        \textbf{DSA-OpenSSL 1.1.1n}
        & 14/4 & 12/4 & 12/4 \\
    \hline
        \textbf{total}
        & 145/41 & 115/41 & 112/41 \\
    \hline
    \end{tabular}
}
\vspace{-5pt}
\end{table}

\rev{\T~\ref{tab:compiler} shows that optimizations affect the analysis results,
as heavy optimizations tend to ``condense'' code into fewer cache lines. Similar
to \T~\ref{tab:resultsoverview}, we provide the discovered leakage sites as well
as grouped leakage units. \tool\ can accurately capture the subtle leakage
(without making false positives) with its employed cache layout threat model.
With manual efforts, we confirm that \textit{all} cases are true positives.
Indeed, we report that all \texttt{-O2} findings are subsumed by those of
\texttt{-O0}, and all \texttt{-O3} findings are subsumed by \texttt{-O2}
findings. In contrast, we report that CacheD/CacheS yields \textit{identical}
findings across different optimization settings, meaning that they have a
considerable number of false positives under \texttt{-O2} and \texttt{-O3}.}

%% file: 7_discussion.tex
\section{Discussion and Limitation}
\label{sec:discussion}

\begin{table*}[htbp]
\centering
\caption{Checking the correctness of refinement type system in \tool\ by
comparing with taint analysis. ``FPs'' denotes false positives of taint analysis.
We randomly select 100 cases for each setting for confirmation except
ElGamal/Libgcrypt 1.9.4.}
\label{tab:taint-type}
\begin{threeparttable}
  \setlength{\tabcolsep}{2.5pt}
  \resizebox{0.90\linewidth}{!}{
\begin{tabular}{c|c|c|c|c|c|c|c}
\hline
    \textbf{Algorithms}
    &\textbf{Implementations}
    &\textbf{\begin{tabular}[c]{@{}c@{}}Instructions\\ on the traces\end{tabular}}
    &\textbf{\begin{tabular}[c]{@{}c@{}}Tainted\\instructions\end{tabular}}
    &\textbf{\begin{tabular}[c]{@{}c@{}}Tainted registers\\and memory cells\end{tabular}}
    &\textbf{\begin{tabular}[c]{@{}c@{}}Registers and memory\\cells with $\sdd$ types\end{tabular}}
    &\textbf{\begin{tabular}[c]{@{}c@{}}``Over-tainted'' registers\\and memory cells\end{tabular}}
    &\textbf{\begin{tabular}[c]{@{}c@{}}\#FPs in manually confirmed\\100 ``over-tainted'' cases\end{tabular}}\\
\hline
    \textbf{RSA}
    & OpenSSL 1.0.2f
    & 1,620,404 & 4,127 & 3,271 & 3,165 & 106 & 100 \\
    \textbf{RSA}
    & OpenSSL 1.1.0g
    & 822,151 & 4,092 & 3,568 & 3,452 & 116 & 100 \\
    \textbf{RSA}
    & OpenSSL 1.1.0h
    & 28,874 & 9,672 & 7,939 & 5,977 & 1,962 & 100 \\
    \textbf{RSA}
    & OpenSSL 1.1.1n
    & 1,763,970 & 51,933 & 34,518 & 30,276 & 4,242 & 100 \\
    \textbf{RSA}
    & OpenSSL 3.0.2
    & 1,711,746 & 56,218 & 39,799 & 37,507 & 2,292 & 100 \\
    \textbf{RSA}
    & Libgcrypt 1.6.1
    & 806,410 & 130,141 & 125,648 & 100,493 & 25,155 & 100 \\
    \textbf{RSA}
    & Libgcrypt 1.7.3
    & 837,215 & 140,478 & 133,105 & 107,731 & 25,374 & 100 \\
    \textbf{RSA}
    & Libgcrypt 1.9.4
    & 114,733 & 102,132 & 104,536 & 103,676 & 860 & 100 \\
\hline
    \textbf{ElGamal}    
    & Libgcrypt 1.6.1
    & 573,242 & 334,024 & 286,095 & 184,976 & 101,119 & 100 \\
    \textbf{ElGamal}    
    & Libgcrypt 1.7.3
    & 573,866 & 334,194 & 286,213 & 185,297 & 100,916 & 100 \\
    \textbf{ElGamal}    
    & Libgcrypt 1.9.4
    & 4,676 & 1,274 & 1,145 & 1,140 & 5 & 5 \\
\hline
    \textbf{ECDSA}
    & OpenSSL 1.0.1e
    & 2,277,459 & 258,261 & 241,326 & 237,381 & 3,945 & 100 \\
    \textbf{ECDSA}
    & OpenSSL 1.1.0g
    & 415,415 & 140,596 & 124,907 & 110,488 & 14,419 & 100 \\
    \textbf{ECDSA}
    & OpenSSL 1.1.0i
    & 298,463 & 81,988 & 72,090 & 55,810 & 16,280 & 100 \\
    \textbf{ECDSA}
    & OpenSSL 1.1.1n
    & 182,745 & 315 & 220 & 120 & 100 & 100 \\
    \textbf{ECDSA}
    & OpenSSL 3.0.2
    & 164,613 & 315 & 220 & 120 & 100 & 100 \\
\hline
    \textbf{DSA}$\natural$
    & OpenSSL 1.1.0i
    & 18,516 & 4,766 & 3,970 & 3,371 & 599 & 100 \\
    \textbf{DSA(swapped)}$\natural$
    & OpenSSL 1.1.0i
    & 18,608 & 4,698 & 3,899 & 3,230 & 669 & 100 \\
    \textbf{DSA}
    & OpenSSL 1.1.1n
    & 1678 & 578 & 435 & 302 & 133 & 100 \\
    \textbf{DSA}
    & OpenSSL 3.0.2
    & 1678 & 578 & 435 & 302 & 133 & 100 \\
\hline
    \textbf{total}
    &
    & 12,236,462 & 1,660,380 & 1,473,339 & 1,174,814 & 298,525 & 1,905 \\
\hline
\end{tabular}
}
\begin{tablenotes}
\footnotesize
\item $\natural$ DSA (OpenSSL-1.1.0i) and its swapped patch are only evaluated
for the key blinding part.
\end{tablenotes}
\end{threeparttable}
\end{table*}

\parh{Type System Benchmarking.}~\rev{Scientifically, 
it would be ideal to benchmark our refinement type system against some ``synthetic datasets'' to determine their algorithmic
effectiveness and efficiency before evaluating side channel detections, which is a
``downstream'' application of our type system. Nevertheless, it is practically
hard to find a proper (synthetic) dataset to solely evaluate the type system, and using downstream applications to reflect the effectiveness of a type
system is a common evaluation plan used by relevant
works~\cite{DBLP:conf/esop/TomanSSI020,DBLP:conf/icfp/VazouSJVJ14,DBLP:conf/pldi/VekrisCJ16}.
To avoid potential confusion, we revisit the effectiveness
and efficiency of our type system as follows.}

\rev{First, our type system is sound (per Proposition~\ref{prop:typerule-new}). All
typing rules are intuitive, and there are no ``tricky'' ones implemented
in \tool. Thus, the soundness is at ease. Second, in terms of
efficiency, our implementation manifests approximately $O(n)$ complexity, where $n$ is the number of instructions in a given trace.
\tool\ is
empirically very efficient. As demonstrated in \F~\ref{fig:rsaperformance},
\tool\ manifests a mostly linear growth in terms of the trace length and
processing time. Overall, the end-to-end evaluation on side channel analysis
illustrates the accuracy of \tool, thereby reflecting the effectiveness of its
underlying type systems at large.}

\rev{Further to the above discussion, we empirically evaluate the type system by
comparing it with taint analysis to check correctly-tagged variables. In
general, taint analysis offers a holistic modelling of how secrets propagate
through the program, while our type system is \textit{more precise}. Most taint
analysis implementation is performed at the syntax level (whose cost and
accuracy is conceptually similar to conventional, syntax-level type inference).
In contrast, as shown in \S~\ref{subsec:motivation} and \F~\ref{fig:motivation},
\tool's type system tracks bit-level values/secrets uniformly using refined
types; thus, the type system captures stronger semantics properties, e.g., it
models how blinding obscures secrets. Therefore, properly masked secrets are not
treated as secrets in \tool\ (i.e., they do not have an $\sdd$ type), but taint
analysis will ``over-taint'' them.}

\rev{Recall \tool\ first conducts taint analysis over the Pin-logged trace before performing
type inference. Thus, we compare the number of tainted registers/memory cells
with the number of variables of type $\sdd$ over the same trace. \T~\ref{tab:taint-type} reports the evaluation results.
As clarified above and observed in \T~\ref{tab:taint-type}, the
number of variables of $\sdd$ type is less than the number of tainted
variables, as expected. Also, we
confirm that \textit{all} variables of type $\sdd$ exist in the tainted set, i.e., our type inference phase has no false negatives (when using tainted
variables as the baseline). 
More importantly, we also manually study every ``over-tainted'' variable that does
not have type $\sdd$. As shown in the 7th column of \T~\ref{tab:taint-type},
taint analysis finds considerably more tainted variables than type inference.
Given the difficulty of manual inspection, for each evaluation setting, we
randomly select 100 cases (if there are more than 100 cases). For each case, we
comprehend the causality of how variable is tainted, and decide if this is
a true positive (meaning that the tainted variable is carrying secrets
correctly) or not.}

\rev{We show the manual inspection results in the last column of
\T~\ref{tab:taint-type}. We find that all the ``over-tainted'' variables are
\textit{false positives} of the taint analysis. It is thus correct for our
type system to neglect them. Among in total 1,905 randomly selected
cases, the ``over-tainted'' variables belong to the following categories:
\ding{172} variables of $\sdd$ type that have been appropriately masked
with blinding, while they are still tainted, \ding{173} variables that 
are further tainted by variables belonging to \ding{172},
\ding{174} variables of $\sdd$ type that have been zeroized by constants,
whereas taint analysis retains the taint label over those variables, and
\ding{175} the base address of a secret buffer is deemed as a taint source, such
that whenever loading from the base address, the output will be tainted. 
While \ding{172}, \ding{173}, and \ding{174} are due to the inherent limitation
of standard taint analysis technique, \ding{175} is due to the ``clumsy''
implementation of our adopted taint analysis tool.\footnote{\rev{We use the taint
analysis tool provided by CacheD. Note that \ding{175} eases the implementation
of a taint engine, but overestimates secrets. Secrets (and their associated
non-secret data) are often stored in a BIGNUM struct. By treating the base
address of this struct as the taint source, non-secret data in the struct are
all tainted due to \ding{175}.}} Out of 1,905 manually checked cases, we find 
that about 52\% cases fall in \ding{175}, whereas the remaining 48\% cases are
due to \ding{172}, \ding{173}, or \ding{174}. Thus, we estimate that around 143K
($298525 \times 48\%$) false positives are due to the inherent limitation of taint
analysis, which are correctly eliminated by our refinement type system.}

\noindent \parh{Extension.}~\rev{We discuss the extension of \tool\ from both
architectural and analysis target perspectives. First, the current
implementation of \tool\ supports to analyze 32-bit x86 binaries. Given that the
closely-related works (e.g., CacheD, CacheS, and CacheAudit) only support 32-bit
x86 binaries, supporting the same binary format enables an ``apple-to-apple''
comparison. Moreover, \tool\ can be extended to 64-bit binaries with no extra
research challenge. We expect to convert each refinement type, currently a
32-bit vector, to a 64-bit vector. We also need to handle new instructions.
Nevertheless, these are engineering endeavors rather than open-ended research
problems. We leave it as one future work to support other architectures
including 64-bit x86.}

\rev{Also, from the analysis target perspective, side channel analyzers in this field
require to flag program secrets (or other sensitive data) specified by users,
and then start to analyze their influence on cache. Detectors (including \tool)
are \textit{not} limited to crypto software. Analyzing crypto software targeted
by previous analyzers, however, makes it easier to compare \tool\ with them.
Given the scalability of \tool, it should be feasible to extend \tool\ to
analyze production software running in trusted execution environments (TEEs) and
detect their side channel
leaks~\cite{ahmad2019obfuscuro,chen2019sgxpectre,wang2019time}.}

%% file: 8_relatedwork.tex
\section{Related Work}
\label{sec:related}

Perfect masking analysis conducted on power side channels is highly relevant to
our work~\cite{kocher1999differential,moradi2011vulnerability}. In such
analysis, all intermediate computation outputs are statistically examined for
independence between secret data and power side channels. 
Recent efforts employ a type-based technique to deduce potentially leakage of
program intermediate variables. Specifically,~\cite{barthe2015verified,
barthe2016strong, el2017symbolic} use a syntactic type system that primarily
relies on the variable structural information.~\cite{zhang2018sc,
gao2019verifying} extend the syntax-based approach to a semantic-based type
system that refines inference rules for boolean masking scheme analysis. Two
improvements~\cite{gao2019quantitative, pengfei2020formal} add rules for
additive and multiplicative masking. These works inspire the design of our
refinement type system. However, crucial gaps exist in applying these rules to
detect cache side channels. First, perfect masking analysis of software power
side channel countermeasures targets specific masked programs (often bitwise
operations), whose computation is usually straightforward (calculating and then
assigning). 
Cache side channel analysis targets complicated production cryptosystems. Type
systems proposed in prior works are primarily for bitvector logical operations,
not general x86 assembly semantics. Second, our tentative exploration shows that
earlier typing rules were often incomplete; they may need to use constraint
solving when typing rules cannot be applied. Their performance is therefore
downgraded. In contrast, \tool's type inference rules completely infer refined
types for variables.

%% file: 9_conclusion.tex
\section{Conclusions}
\label{sec:conclusion}

Detecting cache side channels in production cryptographic software is still an
open problem. This paper presents \tool, a refinement type-based tool to deliver
highly efficient and accurate analysis of cache side channels over x86 binary
code. Evaluation over real-world cryptographic software shows that \tool\
identifies side channels with high precision, efficiency, and scalability.

%% file: 10_appendices.tex
\section{One-bitvector Constant Type Rules for Logical Operations}
\label{appdix:rules}

\F~\ref{fig:bitrule-constant} shows the one-bitvector type rules involving the $\cst$ type. Rule \textsc{Const-Conj} and \textsc{Cons-Disj} rules handle the situation when the refined type of one operand expression is $\cst$. These four rules are straightforward. Rule \textsc{XOR.III} and \textsc{XOR.IV} describe the refined type $\cst$ cases, which are consistent with basic cognition, and the value predicates are given. Rules \textsc{Neg.II} and \textsc{Neg.III} keep security types unchanged while tracking values precisely in types.

\section{Type Rules for Statements}
\label{appedix:rules_stmt}

We extend the type environment $\Gamma$ (defined in \S~\ref{bitlevel}) to track the value and security type of each element in a vector, i.e., $\Gamma ::= \emptyset \mid \Gamma, x: \rho \mid \Gamma, e_1[e_2] : \rho$. We use the following rule to derive the type of vector indexing expression:

\vspace{-10pt}
\begin{mathpar}
	\inferrule*[Left = T-Vec-Index] 
	{ e_1[e_2] \in \Gamma }
	{ \Gamma \vdash e_1[e_2] : \rho} 
\end{mathpar}

\F~\ref{fig:typerule-stmt} shows the type rules for statements.
It tracks values in a flow-sensitive way.
The type judgment is in the form $\Gamma \vdash S \dashv \Gamma'$, meaning that statement $S$ is type-checked under $\Gamma$, and produces a new type environment $\Gamma'$.
The notation $\Gamma[x \mapsto \rho]$ overrides $x$'s type with $\rho$ in $\Gamma$ if $x$ is in $\Gamma$; otherwise extends $\Gamma$ with $[x \mapsto \rho]$.

Rules \textsc{Assign-I} and \textsc{Assign-II} are for assignments, with and without the presence of a value predicate respectively. Rule \textsc{Assign-I} updates variable $x$'s value predicate with the one on the right-hand side, enabling precise tracking values in types.
Rules \textsc{Load} and \textsc{Store} are used for reading from and writing into memories, where memory access safety is assumed.
Rules \textsc{Load-I} and \textsc{Load-II} retrieve the type of the vector $e_1$ at index $e_2$ and update $x$'s type with it if $e_1[e_2]$ is already tracked in $\Gamma$. Otherwise, $x$'s type is updated with $\tau_1 \sqcup \tau_2$ (rule \textsc{Load-III}), which is capable of tracking implicit information flow.
Rules \textsc{Store-I} and \textsc{Store-II} update the type of the element $e_1[e_2]$ with $x$'s type. 
Rule \textsc{Seq} checks the first instruction $S_1$ under $\Gamma$ and produces a new type environment $\Gamma''$, under which, instruction $S_2$ is checked.

\begin{figure}[htp]
\footnotesize
\begin{mathpar}
\inferrule[Prim-I]{}{\Gamma \vdash 0: \{v:B \mid v = 0 \land v: \cst \}}
\and
\inferrule[Prim-II]{}{\Gamma \vdash 1: \{v:B \mid v = 1 \land v: \cst \}}\\
\and
\inferrule[Const-Conj.I]{\Gamma \vdash e_1: \{v:B \mid v: \tau_1 \} \\\\ \Gamma \vdash e_2: \{v:B \mid v = 0 \land v: \cst \}}
{\Gamma \vdash e_1 \land e_2: \{v:B \mid v = 0 \land v: \cst \}}
\and
\inferrule[Const-Conj.II]{\Gamma \vdash e_1: \{v:B \mid v: \tau_1 \} \\\\ \Gamma \vdash e_2: \{v: B \mid v = 1 \land v: \cst \}} 
{\Gamma \vdash e_1 \land e_2: \{v:B \mid v: \tau_1\}}\\
\and
\inferrule[Const-Disj.I]{\Gamma \vdash e_1: \{v: B \mid v: \tau_1\} \\\\ \Gamma \vdash e_2: \{v:B \mid v = 0 \land v: \cst \}}
{\Gamma \vdash e_1 \lor e_2: \{v: B \mid v: \tau_1 \}}
\and
\inferrule[Const-Disj.II]{\Gamma \vdash e_1: \{v: B \mid \tau_1\} \\\\ \Gamma \vdash e_2: \{v: B \mid v = 1 \land v: \tau_2 \}} 
{\Gamma \vdash e_1 \lor e_2: \{v: B \mid v = 1 \land v: \cst \}}\\
\and
\inferrule[XOR.III]{\Gamma \vdash e_1: \{v: B \mid v: \cst \} \\\\ \Gamma \vdash e_2: \{v: B \mid v: \cst\} \\ e_1 \neq e_2 }
{\Gamma \vdash e_1 \oplus e_2: \{v: B \mid v = 1 \land v: \cst\}}
\and
\inferrule[XOR.IV]{\Gamma \vdash e: \{v: B \mid v: \cst \}}
{\Gamma \vdash e \oplus e: \{v:B \mid v = 0 \land v: \cst\}}\\
\and
\inferrule[Neg.II]{\Gamma \vdash e: \{v:B \mid v = 0 \land v: \cst \}} 
{\Gamma \vdash \neg e: \{v:B \mid v = 1 \land v: \cst \}}
\and
\inferrule[Neg.III]{\Gamma \vdash e: \{v:B \mid v = 1 \land v: \cst \}} 
{\Gamma \vdash \neg e: \{v: B \mid v = 0 \land v: \cst \}}
\end{mathpar}
\caption{One-bitvector Constant Type Rules.}
\vspace{-5pt}
\label{fig:bitrule-constant}
\end{figure}

\begin{figure}[htp]
\footnotesize
\begin{mathpar}
\inferrule[Assign-I]
{ \Gamma \vdash x : \{ v: B \mid v: \tau_x \} \\ \Gamma \vdash e : \{v: B \mid v = b \land v: \tau_e \} \\ \Gamma' = \Gamma[x \mapsto \{v: B \mid v = b \land v: \tau_e \} ] }
{ \Gamma \vdash x \leftarrow e \dashv \Gamma'}
\and 
\inferrule[Assign-II]
{ \Gamma \vdash x : \{ v: B \mid v: \tau_x \} \\ \Gamma \vdash e : \{v: B \mid v: \tau_e \} \\ \Gamma' = \Gamma[x \mapsto \{v: B \mid v: \tau_e \}] }
{ \Gamma \vdash x \leftarrow e \dashv \Gamma'}
\and \vspace{-10pt}
\inferrule[Load-I]
{ \Gamma \vdash x : \{ v: B \mid v: \tau_x \} \\ \Gamma \vdash e_1[e_2]: \{v : B \mid v = b \land v : \tau_v\} \\ \Gamma' = \Gamma[x \mapsto \{v: B \mid v = b \land v: \tau_v\}] }
{ \Gamma \vdash x \leftarrow e_1[e_2] \dashv \Gamma'}
\and
\inferrule[Load-II]
{ \Gamma \vdash x : \{ v: B \mid v: \tau_x \} \\ \Gamma \vdash e_1[e_2]: \{v : B \mid v : \tau_v\} \\ \Gamma' = \Gamma[x \mapsto \{v: B \mid v: \tau_v\}] }
{ \Gamma \vdash x \leftarrow e_1[e_2] \dashv \Gamma'}
\and
\inferrule[Load-III]
{ \Gamma \vdash x : \{ v: B \mid v: \tau_x \} \\ \Gamma \vdash e_1 : \{v: \bv{n} \mid v: \tau_1 \} \\ \Gamma \vdash e_2 : \{v: \bv{n} \mid v: \tau_2 \} \\ \Gamma \not\vdash e_1[e_2] \\ \tau_r = \tau_1 \sqcup \tau_2 \\ \Gamma' = \Gamma[x \mapsto \{v: B \mid v : \tau_r \} }
{ \Gamma \vdash x \leftarrow e_1[e_2] \dashv \Gamma'}
\and
\inferrule[Store-I]
{ \Gamma \vdash e_1: \{v: \bv{n} \mid v: \tau_1 \} \\ \Gamma \vdash e_2 : \{v: \bv{n} \mid v: \tau_2 \} \\ \Gamma \vdash x : \{ v: B \mid v = b \land v: \tau_x \} \\ \Gamma' = \Gamma[e_1[e_2] : \{v: B \mid v = b \land v: \tau_x \}] }
{ \Gamma \vdash e_1[e_2] \leftarrow x \dashv \Gamma'}
\and
\inferrule[Store-II]
{ \Gamma \vdash e_1: \{v: \bv{n} \mid v: \tau_1 \} \\ \Gamma \vdash e_2 : \{v: \bv{n} \mid v: \tau_2 \} \\ \Gamma \vdash x : \{ v: B \mid v: \tau_x \} \\ \Gamma' = \Gamma[e_1[e_2] : \{v: B \mid v: \tau_x \}] }
{ \Gamma \vdash e_1[e_2] \leftarrow x \dashv \Gamma'}
\and
\inferrule[Seq]
{\Gamma \vdash s_1 \dashv \Gamma'' \\ \Gamma'' \vdash s_2 \dashv \Gamma'}
{\Gamma \vdash s_1;s_2 \dashv \Gamma'}
\end{mathpar}
\caption{Type Rules for Statements.}
\vspace{-5pt}
\label{fig:typerule-stmt}
\end{figure}

\begin{figure}[htp]
	\centering
	\footnotesize
	\begin{tabular}{ll}
		&$1\ void\ \_gcry\_mpi\_powm(gcry\_mpi\_t\ res,\ gcry\_mpi\_t\ base,$\\
		&$2\ \quad gcry\_mpi\_t\ expo,\ gcry\_mpi\_t\ mod)\lbrace$\\
		&$3\ \quad \cdots$\\ 
		&$4\ \quad \textcolor{red}{e} = ep[i];$\\
		&$5\ \quad count\_leading\_zeros(\textcolor{red}{c},\ \textcolor{red}{e});$\\
		&$6\ \quad \textcolor{red}{e} = (\textcolor{red}{e} \ll \textcolor{red}{c}) \ll 1;$\\
		&$7\ \quad \cdots$\\
		&$8\ \quad \textcolor{red}{e0} = (\textcolor{red}{e} \gg (BITS\_PER\_MPI\_LIMB - W));$\\
		&$9\ \quad count\_trailing\_zeros(\textcolor{red}{c0},\ \textcolor{red}{e0});$\\
		&$10\ \quad \textcolor{red}{e0} = (\textcolor{red}{e0} \gg \textcolor{red}{c0}) \gg 1;$\\
		&$11\ \quad \cdots$\\
		&$12\ \quad base\_u = b\_2i3[\textcolor{red}{e0} - 1];$\\
		&$13\ \quad base\_u\_size = b\_2i3size[\textcolor{red}{e0} - 1];$\\
		&$14\ \quad \cdots$\\
		&$15\ \rbrace$
	\end{tabular}
	\caption{RSA/Elgamal information leaks found in Libgcrypt-1.6.1.}
	\label{fig:vulknownlibg}
	\vspace{-10pt}
\end{figure}

\begin{figure}[htp]
	\centering
	\footnotesize
	\begin{tabular}{ll}
		&$1\ int\ BN\_num\_bits(const\ BIGNUM\ \ast a) \lbrace$\\
		&$2\ \quad int\ i = a -> top - 1;$\\
		&$3\ \quad bn\_check\_top(a);$\\
		&$4\ \quad if\ (BN\_is\_zero(a))\ return\ 0;$\\
		&$5\ \quad return\ ((i \ast BN\_BITS2) + BN\_num\_bits\_word(a -> d[i]));$\\
		&$6\ \rbrace$\\
		&$7\ int\ BN\_num\_bits\_word(BN\_ULONG\ \textcolor{red}{l}) \lbrace$\\
		&$8\ \quad static\ const\ char\ bits[256] = \lbrace$\\
		&$9\ \quad \quad 0,1,2,2,3,3,3,3,4,4,4,4,4,4,4,4,$\\
		&$10\ \quad \quad \cdots$\\
		&$11\ \quad \quad 8,8,8,8,8,8,8,8,8,8,8,8,8,8,8,8,$\\
		&$12\ \quad \rbrace;$\\
		&$13\ \quad if\ (\textcolor{red}{l}\ \&\ 0xffff0000L) \lbrace$\\
		&$14\ \quad \quad if\ (\textcolor{red}{l}\ \&\ 0xff000000L)\ return\ bits[\textcolor{red}{l} >> 24] + 24;$\\
		&$15\ \quad \quad else\ return\ bits[\textcolor{red}{l} >> 16] + 16;$\\
		&$16\ \quad \rbrace$\\
		&$17\ \quad else $\\
		&$18\ \quad \quad if\ (\textcolor{red}{l}\ \&\ 0xff00L)\ return\ bits[\textcolor{red}{l} \gg 8] + 8;$\\
		&$19\ \quad \quad else\ return\ bits[\textcolor{red}{l}];$\\
		&$20\ \quad \rbrace$\\
		&$21\ \rbrace$\\
	\end{tabular}
	\caption{RSA information leaks found in OpenSSL-1.0.2f.}
	\vspace{-10pt}
	\label{fig:vulknownrsa}
\end{figure}

\section{Known Information Leaks}
\label{appdix:known}

\noindent \textbf{RSA/Elgamal-Libgcrypt.}~\F~\ref{fig:vulknownlibg} demonstrates the sliding-window implementation. Before the sliding-window algorithm, two lookup tables are constructed. The first table stores the modular exponentiation values of various bases and the second one stores the length of the corresponding value. In the main loop of modular exponentiation, symbol \texttt{e} represents the element of secret array and \texttt{e0} represents each sliding-window of \texttt{e}. Then \texttt{e0} is used to access the pre-computation tables. Intuitively, with a $\textsc{PRIME-PROBE}$ attack, different cache sets are observed accessed under different sliding-window values, eventually leaking the secret \texttt{e}.

\noindent \textbf{RSA-OpenSSL.}~The \textrm{BN\_num\_bits\_word} is called by \textrm{BN\_num\_bits} that
counts the number of bits of a secret (see \F~\ref{fig:vulknownrsa}). The
secret is stored in a \textrm{BIGNUM} struct, where the key value is stored
in a byte array \texttt{a->d} of 32-bit element, and the length of the
array is stored in \texttt{a->top}. The number of bits of the last element
requires determined separately because its valid bits may be less than 32 bits.
Therefore, function \textrm{BN\_num\_bits\_word} refers to a lookup table to
determine the number of bits of the last element of the secret array. Different
last elements lead to different entries of the lookup table being accessed.
Meanwhile, the branches further narrow down the secret length. Thus, it is a
combined vulnerability of memory access and branch.

\noindent \textbf{ECDSA-OpenSSL.}~ECDSA performs the second step of signature as $s = (r \cdot priv\_key + m)\ mod\ order$ (see \F~\ref{fig:vulknownecdsa}), where $r \cdot priv\_key$ represents the result of the first step that multiplies part of the signature $r$ with the private key $priv\_key$. Before the addition operation of the second step, the value of $r \cdot priv\_key$ ensures to be reduced into the range [0, $order$-1]. By observing whether a reduction behaves after the addition operation inside the second step (function \textrm{BN\_mod\_add\_quick}), an attacker can deduce the range information of the private key $priv\_key$.

\begin{figure}[htb]
	\centering
	\footnotesize
	\begin{tabular}{ll}
		&$1\ ECDSA\_SIG\ \ast ossl\_ecdsa\_sign\_sig(\cdots) \lbrace$\\
		&$2\ \quad \cdots$\\
		&$3\ \quad do \lbrace$\\
		&$4\ \quad \quad \cdots$\\
		&$5\ \quad \quad if\ (!BN\_mod\_mul(tmp,\ priv\_key,\ ret->r,\ order,\ ctx)) \lbrace$\\
		&$6\ \quad \quad \quad ECerr(EC\_F\_OSSL\_ECDSA\_SIGN\_SIG,\ ERR\_R\_BN\_LIB);$\\
		&$7\ \quad \quad \quad goto\ err;$\\
		&$8\ \quad \quad \rbrace$\\
		&$9\ \quad \quad if\ (!BN\_mod\_add\_quick(s,\ tmp,\ m,\ order)) \lbrace$\\
		&$10\ \quad \quad \quad ECerr(EC\_F\_OSSL\_ECDSA\_SIGN\_SIG,\ ERR\_R\_BN\_LIB);$\\
		&$11\ \quad \quad \quad goto\ err;$\\
		&$12\ \quad \quad \rbrace$\\
		&$13\ \quad \quad \cdots$\\
		&$14\ \quad \rbrace $\\
		&$15\ \quad while(1);$\\
		&$16\ \quad \cdots$\\
		&$17\ \rbrace$\\
		&$18\ int\ BN\_mod\_add\_quick(BIGNUM \ast r,$\\
		&$19\ \quad const\ BIGNUM \ast \textcolor{red}{a},\ const\ BIGNUM \ast b,$\\
		&$20\ \quad const\ BIGNUM \ast m) \lbrace$\\
		&$21\ \quad if\ (!BN\_uadd(r,\ \textcolor{red}{a},\ b))$\\
		&$22\ \quad \quad return\ 0;$\\
		&$23\ \quad if\ (!BN\_ucmp(\textcolor{red}{r},\ m) >= 0)$\\
		&$24\ \quad \quad return\ BN\_usub(\textcolor{red}{r},\ \textcolor{red}{r},\ m);$\\
		&$25\ \quad return\ 1;$\\
		&$26\ \rbrace$\\
	\end{tabular}
	\caption{ECDSA information leaks found in OpenSSL-1.1.0g.}
	\label{fig:vulknownecdsa}
\end{figure}

\section{Unknown Information Leaks in OpenSSL}
\label{appdix:unknown}

\begin{figure}[htbp]
	\centering
	\footnotesize
	\begin{tabular}{ll}
		&$1\ int\ BN\_rshift1(BIGNUM \ast r,\ const\ BIGNUM \ast a) \lbrace$\\
		&$2\ \quad \cdots$\\
		&$3\ \quad \textcolor{red}{i} = a->top;$\\
        &$4\ \quad ap = a->d;$\\
        &$5\ \quad \cdots$\\
        &$6\ \quad rp = r->d;$\\
        &$7\ \quad t = ap[--\textcolor{red}{i}];$\\
        &$8\ \quad c = (t \& 1) ?\ BN\_TBIT : 0;$\\
		&$9\ \quad if\ (t \gg= 1)$\\
		&$10\ \quad \quad rp[\textcolor{red}{i}] = t;$\\
		&$11\ \quad while\ (\textcolor{red}{i} > 0) \lbrace$\\		
		&$12\ \quad \quad t = ap[--\textcolor{red}{i}];$\\
		&$13\ \quad \quad rp[\textcolor{red}{i}] = ((t \gg 1) \& BN\_MASK2) \mid c;$\\
		&$14\ \quad \quad c = (t \& 1) ?\ BN\_TBIT : 0;$\\
		&$15\ \quad \rbrace$\\
		&$16\ \quad \cdots$\\
		&$17\ \rbrace$\\
	\end{tabular}
	\caption{BN\_rshift1 information leaks found in OpenSSL-1.1.0g.}
	\label{fig:vulunknown1}
\end{figure}

\begin{figure}[htbp]
	\centering
	\footnotesize
	\begin{tabular}{ll}
		&$1\ int\ bn\_mul\_normal(BN\_ULONG \ast r,$\\
		&$2\ \quad BN\_ULONG \ast a,\ int\ na,\ BN\_ULONG \ast b,\ int\ \textcolor{red}{nb}) \lbrace$\\
		&$3\ \quad \cdots$\\
		&$4\ \quad for(;;) \lbrace$\\
        &$5\ \quad \quad if\ (--\textcolor{red}{nb} <= 0)\ return;$\\
        &$6\ \quad \quad rr[1] = bn\_mul\_add\_words(\&(r[1]), a, na, b[1]);$\\
        &$7\ \quad \quad if\ (--\textcolor{red}{nb} <= 0)\ return;$\\
		&$8\ \quad \quad rr[2] = bn\_mul\_add\_words(\&(r[2]), a, na, b[2]);$\\
		&$9\ \quad \quad if\ (--\textcolor{red}{nb} <= 0)\ return;$\\
		&$10\ \quad \quad rr[3] = bn\_mul\_add\_words(\&(r[3]), a, na, b[3]);$\\
		&$11\ \quad \quad if\ (--\textcolor{red}{nb} <= 0)\ return;$\\
		&$12\ \quad \quad rr[4] = bn\_mul\_add\_words(\&(r[4]), a, na, b[4]);$\\
		&$13\ \quad \quad \cdots$\\
		&$14\ \quad \rbrace$\\
		&$15\ \rbrace$\\
	\end{tabular}
	\caption{bn\_mul\_normal information leaks found in OpenSSL-1.1.0i.}
	\label{fig:vulunknown2}
\end{figure}

%% file: main.bbl

\begin{thebibliography}{73}


\ifx \showCODEN    \undefined \def \showCODEN     #1{\unskip}     \fi
\ifx \showDOI      \undefined \def \showDOI       #1{#1}\fi
\ifx \showISBNx    \undefined \def \showISBNx     #1{\unskip}     \fi
\ifx \showISBNxiii \undefined \def \showISBNxiii  #1{\unskip}     \fi
\ifx \showISSN     \undefined \def \showISSN      #1{\unskip}     \fi
\ifx \showLCCN     \undefined \def \showLCCN      #1{\unskip}     \fi
\ifx \shownote     \undefined \def \shownote      #1{#1}          \fi
\ifx \showarticletitle \undefined \def \showarticletitle #1{#1}   \fi
\ifx \showURL      \undefined \def \showURL       {\relax}        \fi
\providecommand\bibfield[2]{#2}
\providecommand\bibinfo[2]{#2}
\providecommand\natexlab[1]{#1}
\providecommand\showeprint[2][]{arXiv:#2}

\bibitem[Ahmad et~al\mbox{.}(2019)]%
        {ahmad2019obfuscuro}
\bibfield{author}{\bibinfo{person}{Adil Ahmad}, \bibinfo{person}{Byunggill
  Joe}, \bibinfo{person}{Yuan Xiao}, \bibinfo{person}{Yinqian Zhang},
  \bibinfo{person}{Insik Shin}, {and} \bibinfo{person}{Byoungyoung Lee}.}
  \bibinfo{year}{2019}\natexlab{}.
\newblock \showarticletitle{Obfuscuro: A commodity obfuscation engine on intel
  sgx}. In \bibinfo{booktitle}{\emph{Network and Distributed System Security
  Symposium}}.
\newblock


\bibitem[Aranha et~al\mbox{.}(2020)]%
        {aranha2020ladderleak}
\bibfield{author}{\bibinfo{person}{Diego~F Aranha},
  \bibinfo{person}{Felipe~Rodrigues Novaes}, \bibinfo{person}{Akira Takahashi},
  \bibinfo{person}{Mehdi Tibouchi}, {and} \bibinfo{person}{Yuval Yarom}.}
  \bibinfo{year}{2020}\natexlab{}.
\newblock \showarticletitle{Ladderleak: Breaking ecdsa with less than one bit
  of nonce leakage}. In \bibinfo{booktitle}{\emph{Proceedings of the 2020 ACM
  SIGSAC Conference on Computer and Communications Security}}.
  \bibinfo{pages}{225--242}.
\newblock


\bibitem[Bao et~al\mbox{.}(2021)]%
        {bao2021abacus}
\bibfield{author}{\bibinfo{person}{Qinkun Bao}, \bibinfo{person}{Zihao Wang},
  \bibinfo{person}{Xiaoting Li}, \bibinfo{person}{James~R Larus}, {and}
  \bibinfo{person}{Dinghao Wu}.} \bibinfo{year}{2021}\natexlab{}.
\newblock \showarticletitle{Abacus: Precise side-channel analysis}. In
  \bibinfo{booktitle}{\emph{2021 IEEE/ACM 43rd International Conference on
  Software Engineering (ICSE)}}. IEEE, \bibinfo{pages}{797--809}.
\newblock


\bibitem[Barthe et~al\mbox{.}(2015)]%
        {barthe2015verified}
\bibfield{author}{\bibinfo{person}{Gilles Barthe}, \bibinfo{person}{Sonia
  Bela{\"\i}d}, \bibinfo{person}{Fran{\c{c}}ois Dupressoir},
  \bibinfo{person}{Pierre-Alain Fouque}, \bibinfo{person}{Benjamin
  Gr{\'e}goire}, {and} \bibinfo{person}{Pierre-Yves Strub}.}
  \bibinfo{year}{2015}\natexlab{}.
\newblock \showarticletitle{Verified proofs of higher-order masking}. In
  \bibinfo{booktitle}{\emph{Annual International Conference on the Theory and
  Applications of Cryptographic Techniques}}. Springer,
  \bibinfo{pages}{457--485}.
\newblock


\bibitem[Barthe et~al\mbox{.}(2016)]%
        {barthe2016strong}
\bibfield{author}{\bibinfo{person}{Gilles Barthe}, \bibinfo{person}{Sonia
  Bela{\"\i}d}, \bibinfo{person}{Fran{\c{c}}ois Dupressoir},
  \bibinfo{person}{Pierre-Alain Fouque}, \bibinfo{person}{Benjamin
  Gr{\'e}goire}, \bibinfo{person}{Pierre-Yves Strub}, {and}
  \bibinfo{person}{R{\'e}becca Zucchini}.} \bibinfo{year}{2016}\natexlab{}.
\newblock \showarticletitle{Strong non-interference and type-directed
  higher-order masking}. In \bibinfo{booktitle}{\emph{Proceedings of the 2016
  ACM SIGSAC Conference on Computer and Communications Security}}.
  \bibinfo{pages}{116--129}.
\newblock


\bibitem[Barthe et~al\mbox{.}(2014)]%
        {barthe2014probabilistic}
\bibfield{author}{\bibinfo{person}{Gilles Barthe}, \bibinfo{person}{C{\'e}dric
  Fournet}, \bibinfo{person}{Benjamin Gr{\'e}goire},
  \bibinfo{person}{Pierre-Yves Strub}, \bibinfo{person}{Nikhil Swamy}, {and}
  \bibinfo{person}{Santiago Zanella-B{\'e}guelin}.}
  \bibinfo{year}{2014}\natexlab{}.
\newblock \showarticletitle{Probabilistic relational verification for
  cryptographic implementations}.
\newblock \bibinfo{journal}{\emph{ACM SIGPLAN Notices}} \bibinfo{volume}{49},
  \bibinfo{number}{1} (\bibinfo{year}{2014}), \bibinfo{pages}{193--205}.
\newblock


\bibitem[Benger et~al\mbox{.}(2014)]%
        {benger2014ooh}
\bibfield{author}{\bibinfo{person}{Naomi Benger}, \bibinfo{person}{Joop van~de
  Pol}, \bibinfo{person}{Nigel~P Smart}, {and} \bibinfo{person}{Yuval Yarom}.}
  \bibinfo{year}{2014}\natexlab{}.
\newblock \showarticletitle{?Ooh Aah... Just a Little Bit?: a small amount of
  side channel can go a long way}. In \bibinfo{booktitle}{\emph{International
  Workshop on Cryptographic Hardware and Embedded Systems}}. Springer,
  \bibinfo{pages}{75--92}.
\newblock


\bibitem[Bengtson et~al\mbox{.}(2011)]%
        {bengtson2011refinement}
\bibfield{author}{\bibinfo{person}{Jesper Bengtson},
  \bibinfo{person}{Karthikeyan Bhargavan}, \bibinfo{person}{C{\'e}dric
  Fournet}, \bibinfo{person}{Andrew~D Gordon}, {and} \bibinfo{person}{Sergio
  Maffeis}.} \bibinfo{year}{2011}\natexlab{}.
\newblock \showarticletitle{Refinement types for secure implementations}.
\newblock \bibinfo{journal}{\emph{ACM Transactions on Programming Languages and
  Systems (TOPLAS)}} \bibinfo{volume}{33}, \bibinfo{number}{2}
  (\bibinfo{year}{2011}), \bibinfo{pages}{1--45}.
\newblock


\bibitem[Bhargavan et~al\mbox{.}(2010)]%
        {bhargavan2010modular}
\bibfield{author}{\bibinfo{person}{Karthikeyan Bhargavan},
  \bibinfo{person}{C{\'e}dric Fournet}, {and} \bibinfo{person}{Andrew~D
  Gordon}.} \bibinfo{year}{2010}\natexlab{}.
\newblock \showarticletitle{Modular verification of security protocol code by
  typing}.
\newblock \bibinfo{journal}{\emph{ACM Sigplan Notices}} \bibinfo{volume}{45},
  \bibinfo{number}{1} (\bibinfo{year}{2010}), \bibinfo{pages}{445--456}.
\newblock


\bibitem[Bhargavan et~al\mbox{.}(2013)]%
        {bhargavan2013implementing}
\bibfield{author}{\bibinfo{person}{Karthikeyan Bhargavan},
  \bibinfo{person}{C{\'e}dric Fournet}, \bibinfo{person}{Markulf Kohlweiss},
  \bibinfo{person}{Alfredo Pironti}, {and} \bibinfo{person}{Pierre-Yves
  Strub}.} \bibinfo{year}{2013}\natexlab{}.
\newblock \showarticletitle{Implementing TLS with verified cryptographic
  security}. In \bibinfo{booktitle}{\emph{2013 IEEE Symposium on Security and
  Privacy}}. IEEE, \bibinfo{pages}{445--459}.
\newblock


\bibitem[Boneh and Venkatesan(1996)]%
        {boneh1996hardness}
\bibfield{author}{\bibinfo{person}{Dan Boneh} {and}
  \bibinfo{person}{Ramarathnam Venkatesan}.} \bibinfo{year}{1996}\natexlab{}.
\newblock \showarticletitle{Hardness of computing the most significant bits of
  secret keys in Diffie-Hellman and related schemes}. In
  \bibinfo{booktitle}{\emph{Annual International Cryptology Conference}}.
  Springer, \bibinfo{pages}{129--142}.
\newblock


\bibitem[Boneh and Venkatesan(1997)]%
        {boneh1997rounding}
\bibfield{author}{\bibinfo{person}{Dan Boneh} {and}
  \bibinfo{person}{Ramarathnam Venkatesan}.} \bibinfo{year}{1997}\natexlab{}.
\newblock \showarticletitle{Rounding in Lattices and its Cryptographic
  Applications.}. In \bibinfo{booktitle}{\emph{SODA}},
  Vol.~\bibinfo{volume}{1997}. Citeseer, \bibinfo{pages}{675--681}.
\newblock


\bibitem[Brotzman et~al\mbox{.}(2019)]%
        {brotzman2019casym}
\bibfield{author}{\bibinfo{person}{Robert Brotzman}, \bibinfo{person}{Shen
  Liu}, \bibinfo{person}{Danfeng Zhang}, \bibinfo{person}{Gang Tan}, {and}
  \bibinfo{person}{Mahmut Kandemir}.} \bibinfo{year}{2019}\natexlab{}.
\newblock \showarticletitle{CaSym: Cache aware symbolic execution for side
  channel detection and mitigation}. In \bibinfo{booktitle}{\emph{2019 IEEE
  Symposium on Security and Privacy (SP)}}. IEEE, \bibinfo{pages}{505--521}.
\newblock


\bibitem[Brumley et~al\mbox{.}(2011)]%
        {brumley2011bap}
\bibfield{author}{\bibinfo{person}{David Brumley}, \bibinfo{person}{Ivan
  Jager}, \bibinfo{person}{Thanassis Avgerinos}, {and}
  \bibinfo{person}{Edward~J Schwartz}.} \bibinfo{year}{2011}\natexlab{}.
\newblock \showarticletitle{BAP: A binary analysis platform}. In
  \bibinfo{booktitle}{\emph{International Conference on Computer Aided
  Verification}}. Springer, \bibinfo{pages}{463--469}.
\newblock


\bibitem[Cardelli(1996)]%
        {cardelli1996type}
\bibfield{author}{\bibinfo{person}{Luca Cardelli}.}
  \bibinfo{year}{1996}\natexlab{}.
\newblock \showarticletitle{Type systems}.
\newblock \bibinfo{journal}{\emph{ACM Computing Surveys (CSUR)}}
  \bibinfo{volume}{28}, \bibinfo{number}{1} (\bibinfo{year}{1996}),
  \bibinfo{pages}{263--264}.
\newblock


\bibitem[Chattopadhyay et~al\mbox{.}(2019)]%
        {chattopadhyay2019quantifying}
\bibfield{author}{\bibinfo{person}{Sudipta Chattopadhyay},
  \bibinfo{person}{Moritz Beck}, \bibinfo{person}{Ahmed Rezine}, {and}
  \bibinfo{person}{Andreas Zeller}.} \bibinfo{year}{2019}\natexlab{}.
\newblock \showarticletitle{Quantifying information leakage in cache attacks
  via symbolic execution}.
\newblock \bibinfo{journal}{\emph{TECS}} (\bibinfo{year}{2019}).
\newblock


\bibitem[Chen et~al\mbox{.}(2019)]%
        {chen2019sgxpectre}
\bibfield{author}{\bibinfo{person}{Guoxing Chen}, \bibinfo{person}{Sanchuan
  Chen}, \bibinfo{person}{Yuan Xiao}, \bibinfo{person}{Yinqian Zhang},
  \bibinfo{person}{Zhiqiang Lin}, {and} \bibinfo{person}{Ten~H Lai}.}
  \bibinfo{year}{2019}\natexlab{}.
\newblock \showarticletitle{Sgxpectre: Stealing intel secrets from sgx enclaves
  via speculative execution}. In \bibinfo{booktitle}{\emph{2019 IEEE European
  Symposium on Security and Privacy (EuroS\&P)}}. IEEE,
  \bibinfo{pages}{142--157}.
\newblock


\bibitem[Cousot and Cousot(1977)]%
        {cousot77abstract}
\bibfield{author}{\bibinfo{person}{P{.} Cousot} {and} \bibinfo{person}{R{.}
  Cousot}.} \bibinfo{year}{1977}\natexlab{}.
\newblock \showarticletitle{Abstract interpretation: a unified lattice model
  for static analysis of programs by construction or approximation of
  fixpoints}. In \bibinfo{booktitle}{\emph{Conference Record of the Fourth
  Annual ACM SIGPLAN-SIGACT Symposium on Principles of Programming Languages}}.
  \bibinfo{pages}{238--252}.
\newblock


\bibitem[Daniel et~al\mbox{.}(2020)]%
        {daniel2020binsec}
\bibfield{author}{\bibinfo{person}{Lesly-Ann Daniel},
  \bibinfo{person}{S{\'e}bastien Bardin}, {and} \bibinfo{person}{Tamara Rezk}.}
  \bibinfo{year}{2020}\natexlab{}.
\newblock \showarticletitle{Binsec/rel: Efficient relational symbolic execution
  for constant-time at binary-level}. In \bibinfo{booktitle}{\emph{2020 IEEE
  Symposium on Security and Privacy (SP)}}. IEEE, \bibinfo{pages}{1021--1038}.
\newblock


\bibitem[Domnitser et~al\mbox{.}(2012)]%
        {domnitser2012non}
\bibfield{author}{\bibinfo{person}{Leonid Domnitser}, \bibinfo{person}{Aamer
  Jaleel}, \bibinfo{person}{Jason Loew}, \bibinfo{person}{Nael Abu-Ghazaleh},
  {and} \bibinfo{person}{Dmitry Ponomarev}.} \bibinfo{year}{2012}\natexlab{}.
\newblock \showarticletitle{Non-monopolizable caches: Low-complexity mitigation
  of cache side channel attacks}.
\newblock \bibinfo{journal}{\emph{ACM Transactions on Architecture and Code
  Optimization (TACO)}} \bibinfo{volume}{8}, \bibinfo{number}{4}
  (\bibinfo{year}{2012}), \bibinfo{pages}{1--21}.
\newblock


\bibitem[Doychev et~al\mbox{.}(2013)]%
        {doychev2013cacheaudit}
\bibfield{author}{\bibinfo{person}{Goran Doychev}, \bibinfo{person}{Dominik
  Feld}, \bibinfo{person}{Boris Kopf}, \bibinfo{person}{Laurent Mauborgne},
  {and} \bibinfo{person}{Jan Reineke}.} \bibinfo{year}{2013}\natexlab{}.
\newblock \showarticletitle{$\{$CacheAudit$\}$: A Tool for the Static Analysis
  of Cache Side Channels}. In \bibinfo{booktitle}{\emph{22nd USENIX Security
  Symposium (USENIX Security 13)}}. \bibinfo{pages}{431--446}.
\newblock


\bibitem[Doychev and K{\"o}pf(2017)]%
        {doychev2017rigorous}
\bibfield{author}{\bibinfo{person}{Goran Doychev} {and} \bibinfo{person}{Boris
  K{\"o}pf}.} \bibinfo{year}{2017}\natexlab{}.
\newblock \showarticletitle{Rigorous analysis of software countermeasures
  against cache attacks}. In \bibinfo{booktitle}{\emph{Proceedings of the 38th
  ACM SIGPLAN Conference on Programming Language Design and Implementation}}.
  \bibinfo{pages}{406--421}.
\newblock


\bibitem[El~Ouahma et~al\mbox{.}(2017)]%
        {el2017symbolic}
\bibfield{author}{\bibinfo{person}{In{\`e}s~Ben El~Ouahma},
  \bibinfo{person}{Quentin~L Meunier}, \bibinfo{person}{Karine Heydemann},
  {and} \bibinfo{person}{Emmanuelle Encrenaz}.}
  \bibinfo{year}{2017}\natexlab{}.
\newblock \showarticletitle{Symbolic approach for side-channel resistance
  analysis of masked assembly codes}. In \bibinfo{booktitle}{\emph{Security
  Proofs for Embedded Systems}}.
\newblock


\bibitem[Eldib et~al\mbox{.}(2014a)]%
        {eldib2014formal}
\bibfield{author}{\bibinfo{person}{Hassan Eldib}, \bibinfo{person}{Chao Wang},
  {and} \bibinfo{person}{Patrick Schaumont}.} \bibinfo{year}{2014}\natexlab{a}.
\newblock \showarticletitle{Formal verification of software countermeasures
  against side-channel attacks}.
\newblock \bibinfo{journal}{\emph{ACM Transactions on Software Engineering and
  Methodology (TOSEM)}} \bibinfo{volume}{24}, \bibinfo{number}{2}
  (\bibinfo{year}{2014}), \bibinfo{pages}{1--24}.
\newblock


\bibitem[Eldib et~al\mbox{.}(2014b)]%
        {eldib2014smt}
\bibfield{author}{\bibinfo{person}{Hassan Eldib}, \bibinfo{person}{Chao Wang},
  {and} \bibinfo{person}{Patrick Schaumont}.} \bibinfo{year}{2014}\natexlab{b}.
\newblock \showarticletitle{SMT-based verification of software countermeasures
  against side-channel attacks}. In \bibinfo{booktitle}{\emph{International
  Conference on Tools and Algorithms for the Construction and Analysis of
  Systems}}. Springer, \bibinfo{pages}{62--77}.
\newblock


\bibitem[Eldib et~al\mbox{.}(2014c)]%
        {eldib2014qms}
\bibfield{author}{\bibinfo{person}{Hassan Eldib}, \bibinfo{person}{Chao Wang},
  \bibinfo{person}{Mostafa Taha}, {and} \bibinfo{person}{Patrick Schaumont}.}
  \bibinfo{year}{2014}\natexlab{c}.
\newblock \showarticletitle{QMS: Evaluating the side-channel resistance of
  masked software from source code}. In \bibinfo{booktitle}{\emph{2014 51st
  ACM/EDAC/IEEE Design Automation Conference (DAC)}}. IEEE,
  \bibinfo{pages}{1--6}.
\newblock


\bibitem[Gao et~al\mbox{.}(2019a)]%
        {gao2019quantitative}
\bibfield{author}{\bibinfo{person}{Pengfei Gao}, \bibinfo{person}{Hongyi Xie},
  \bibinfo{person}{Jun Zhang}, \bibinfo{person}{Fu Song}, {and}
  \bibinfo{person}{Taolue Chen}.} \bibinfo{year}{2019}\natexlab{a}.
\newblock \showarticletitle{Quantitative verification of masked arithmetic
  programs against side-channel attacks}. In
  \bibinfo{booktitle}{\emph{International Conference on Tools and Algorithms
  for the Construction and Analysis of Systems}}. Springer,
  \bibinfo{pages}{155--173}.
\newblock


\bibitem[Gao et~al\mbox{.}(2019b)]%
        {gao2019verifying}
\bibfield{author}{\bibinfo{person}{Pengfei Gao}, \bibinfo{person}{Jun Zhang},
  \bibinfo{person}{Fu Song}, {and} \bibinfo{person}{Chao Wang}.}
  \bibinfo{year}{2019}\natexlab{b}.
\newblock \showarticletitle{Verifying and quantifying side-channel resistance
  of masked software implementations}.
\newblock \bibinfo{journal}{\emph{ACM Transactions on Software Engineering and
  Methodology (TOSEM)}} \bibinfo{volume}{28}, \bibinfo{number}{3}
  (\bibinfo{year}{2019}), \bibinfo{pages}{1--32}.
\newblock


\bibitem[Gruss et~al\mbox{.}(2016)]%
        {gruss2016flush}
\bibfield{author}{\bibinfo{person}{Daniel Gruss},
  \bibinfo{person}{Cl{\'e}mentine Maurice}, \bibinfo{person}{Klaus Wagner},
  {and} \bibinfo{person}{Stefan Mangard}.} \bibinfo{year}{2016}\natexlab{}.
\newblock \showarticletitle{Flush+ Flush: a fast and stealthy cache attack}. In
  \bibinfo{booktitle}{\emph{International Conference on Detection of Intrusions
  and Malware, and Vulnerability Assessment}}. Springer,
  \bibinfo{pages}{279--299}.
\newblock


\bibitem[Jhala et~al\mbox{.}(2021)]%
        {jhala2021refinement}
\bibfield{author}{\bibinfo{person}{Ranjit Jhala}, \bibinfo{person}{Niki Vazou},
  {et~al\mbox{.}}} \bibinfo{year}{2021}\natexlab{}.
\newblock \showarticletitle{Refinement Types: A Tutorial}.
\newblock \bibinfo{journal}{\emph{Foundations and Trends in Programming
  Languages}} \bibinfo{volume}{6}, \bibinfo{number}{3--4}
  (\bibinfo{year}{2021}), \bibinfo{pages}{159--317}.
\newblock


\bibitem[Kocher et~al\mbox{.}(1999)]%
        {kocher1999differential}
\bibfield{author}{\bibinfo{person}{Paul Kocher}, \bibinfo{person}{Joshua
  Jaffe}, {and} \bibinfo{person}{Benjamin Jun}.}
  \bibinfo{year}{1999}\natexlab{}.
\newblock \showarticletitle{Differential power analysis}. In
  \bibinfo{booktitle}{\emph{Annual international cryptology conference}}.
  Springer, \bibinfo{pages}{388--397}.
\newblock


\bibitem[Libressl-1f6b35b(2019)]%
        {libressl@1f6b35b}
\bibfield{author}{\bibinfo{person}{Libressl-1f6b35b}.}
  \bibinfo{year}{2019}\natexlab{}.
\newblock \bibinfo{title}{Remove the blinding later to avoid leaking
  information on the length.}
\newblock
\newblock
\urldef\tempurl%
\url{https://github.com/libressl-portable/openbsd/commit/1f6b35b}
\showURL{%
\tempurl}


\bibitem[Libressl-2cd28f9(2018)]%
        {libressl@2cd28f9}
\bibfield{author}{\bibinfo{person}{Libressl-2cd28f9}.}
  \bibinfo{year}{2018}\natexlab{}.
\newblock \bibinfo{title}{Use a blinding value when generating a DSA
  signature.}
\newblock
\newblock
\urldef\tempurl%
\url{https://github.com/libressl-portable/openbsd/commit/2cd28f9?diff=unified}
\showURL{%
\tempurl}


\bibitem[Liu et~al\mbox{.}(2015)]%
        {liu2015last}
\bibfield{author}{\bibinfo{person}{Fangfei Liu}, \bibinfo{person}{Yuval Yarom},
  \bibinfo{person}{Qian Ge}, \bibinfo{person}{Gernot Heiser}, {and}
  \bibinfo{person}{Ruby~B Lee}.} \bibinfo{year}{2015}\natexlab{}.
\newblock \showarticletitle{Last-level cache side-channel attacks are
  practical}. In \bibinfo{booktitle}{\emph{2015 IEEE symposium on security and
  privacy}}. IEEE, \bibinfo{pages}{605--622}.
\newblock


\bibitem[Lou et~al\mbox{.}(2021)]%
        {lou2021survey}
\bibfield{author}{\bibinfo{person}{Xiaoxuan Lou}, \bibinfo{person}{Tianwei
  Zhang}, \bibinfo{person}{Jun Jiang}, {and} \bibinfo{person}{Yinqian Zhang}.}
  \bibinfo{year}{2021}\natexlab{}.
\newblock \showarticletitle{A Survey of Microarchitectural Side-channel
  Vulnerabilities, Attacks, and Defenses in Cryptography}.
\newblock \bibinfo{journal}{\emph{ACM Computing Surveys (CSUR)}}
  \bibinfo{volume}{54}, \bibinfo{number}{6} (\bibinfo{year}{2021}),
  \bibinfo{pages}{1--37}.
\newblock


\bibitem[Luk et~al\mbox{.}(2005)]%
        {luk2005pin}
\bibfield{author}{\bibinfo{person}{Chi-Keung Luk}, \bibinfo{person}{Robert
  Cohn}, \bibinfo{person}{Robert Muth}, \bibinfo{person}{Harish Patil},
  \bibinfo{person}{Artur Klauser}, \bibinfo{person}{Geoff Lowney},
  \bibinfo{person}{Steven Wallace}, \bibinfo{person}{Vijay~Janapa Reddi}, {and}
  \bibinfo{person}{Kim Hazelwood}.} \bibinfo{year}{2005}\natexlab{}.
\newblock \showarticletitle{Pin: building customized program analysis tools
  with dynamic instrumentation}.
\newblock \bibinfo{journal}{\emph{Acm sigplan notices}} \bibinfo{volume}{40},
  \bibinfo{number}{6} (\bibinfo{year}{2005}), \bibinfo{pages}{190--200}.
\newblock


\bibitem[Moradi et~al\mbox{.}(2011)]%
        {moradi2011vulnerability}
\bibfield{author}{\bibinfo{person}{Amir Moradi}, \bibinfo{person}{Alessandro
  Barenghi}, \bibinfo{person}{Timo Kasper}, {and} \bibinfo{person}{Christof
  Paar}.} \bibinfo{year}{2011}\natexlab{}.
\newblock \showarticletitle{On the vulnerability of FPGA bitstream encryption
  against power analysis attacks: Extracting keys from Xilinx Virtex-II FPGAs}.
  In \bibinfo{booktitle}{\emph{Proceedings of the 18th ACM conference on
  Computer and communications security}}. \bibinfo{pages}{111--124}.
\newblock


\bibitem[Moura and Bj{\o}rner(2008)]%
        {z3}
\bibfield{author}{\bibinfo{person}{Leonardo~De Moura} {and}
  \bibinfo{person}{Nikolaj Bj{\o}rner}.} \bibinfo{year}{2008}\natexlab{}.
\newblock \showarticletitle{{Z3}: an efficient {SMT} solver}
  \emph{(\bibinfo{series}{TACAS})}.
\newblock


\bibitem[Nguyen and Shparlinski(2002)]%
        {nguyen2002insecurity}
\bibfield{author}{\bibinfo{person}{Phong~Q Nguyen} {and}
  \bibinfo{person}{Igor~E Shparlinski}.} \bibinfo{year}{2002}\natexlab{}.
\newblock \showarticletitle{The insecurity of the digital signature algorithm
  with partially known nonces.}
\newblock \bibinfo{journal}{\emph{Journal of Cryptology}} \bibinfo{volume}{15},
  \bibinfo{number}{3} (\bibinfo{year}{2002}).
\newblock


\bibitem[Nguyen and Shparlinski(2003)]%
        {nguyen2003insecurity}
\bibfield{author}{\bibinfo{person}{Phong~Q Nguyen} {and}
  \bibinfo{person}{Igor~E Shparlinski}.} \bibinfo{year}{2003}\natexlab{}.
\newblock \showarticletitle{The insecurity of the elliptic curve digital
  signature algorithm with partially known nonces}.
\newblock \bibinfo{journal}{\emph{Designs, codes and cryptography}}
  \bibinfo{volume}{30}, \bibinfo{number}{2} (\bibinfo{year}{2003}),
  \bibinfo{pages}{201--217}.
\newblock


\bibitem[OpenSSL-2198be3(2014)]%
        {openssl@2198be3}
\bibfield{author}{\bibinfo{person}{OpenSSL-2198be3}.}
  \bibinfo{year}{2014}\natexlab{}.
\newblock \bibinfo{title}{Fix for CVE-2014-0076.}
\newblock
\newblock
\urldef\tempurl%
\url{https://github.com/openssl/openssl/commit/2198be3483259de374f91e57d247d0fc667aef29}
\showURL{%
\tempurl}


\bibitem[OpenSSL-4b7a4ba(2014)]%
        {openssl@4b7a4ba}
\bibfield{author}{\bibinfo{person}{OpenSSL-4b7a4ba}.}
  \bibinfo{year}{2014}\natexlab{}.
\newblock \bibinfo{title}{Fix for CVE-2014-0076.}
\newblock
\newblock
\urldef\tempurl%
\url{https://github.com/openssl/openssl/commit/4b7a4ba29cafa432fc4266fe6e59e60bc1c96332}
\showURL{%
\tempurl}


\bibitem[OpenSSL-972c87d(2018)]%
        {openssl@972c87d}
\bibfield{author}{\bibinfo{person}{OpenSSL-972c87d}.}
  \bibinfo{year}{2018}\natexlab{}.
\newblock \bibinfo{title}{Make bn\_num\_bits\_word constant-time.}
\newblock
\newblock
\urldef\tempurl%
\url{https://github.com/openssl/openssl/commit/972c87dfc7e765bd28a4964519c362f0d3a58ca4}
\showURL{%
\tempurl}


\bibitem[Osvik et~al\mbox{.}(2006)]%
        {osvik2006cache}
\bibfield{author}{\bibinfo{person}{Dag~Arne Osvik}, \bibinfo{person}{Adi
  Shamir}, {and} \bibinfo{person}{Eran Tromer}.}
  \bibinfo{year}{2006}\natexlab{}.
\newblock \showarticletitle{Cache attacks and countermeasures: the case of
  AES}. In \bibinfo{booktitle}{\emph{Cryptographers’ track at the RSA
  conference}}. Springer, \bibinfo{pages}{1--20}.
\newblock


\bibitem[Pengfei et~al\mbox{.}(2020)]%
        {pengfei2020formal}
\bibfield{author}{\bibinfo{person}{Gao Pengfei}, \bibinfo{person}{Xie Hongyi},
  \bibinfo{person}{Pu Sun}, \bibinfo{person}{Jun Zhang}, \bibinfo{person}{Fu
  Song}, {and} \bibinfo{person}{Taolue Chen}.} \bibinfo{year}{2020}\natexlab{}.
\newblock \showarticletitle{Formal verification of masking countermeasures for
  arithmetic programs}.
\newblock \bibinfo{journal}{\emph{IEEE Transactions on Software Engineering}}
  (\bibinfo{year}{2020}).
\newblock


\bibitem[Percival(2005)]%
        {percival2005cache}
\bibfield{author}{\bibinfo{person}{Colin Percival}.}
  \bibinfo{year}{2005}\natexlab{}.
\newblock \bibinfo{title}{Cache missing for fun and profit}.
\newblock
\newblock


\bibitem[Pierce(2002)]%
        {pierce2002types}
\bibfield{author}{\bibinfo{person}{Benjamin~C Pierce}.}
  \bibinfo{year}{2002}\natexlab{}.
\newblock \bibinfo{booktitle}{\emph{Types and programming languages}}.
\newblock \bibinfo{publisher}{MIT press}.
\newblock


\bibitem[Purnal et~al\mbox{.}(2021)]%
        {purnal2021systematic}
\bibfield{author}{\bibinfo{person}{Antoon Purnal}, \bibinfo{person}{Lukas
  Giner}, \bibinfo{person}{Daniel Gruss}, {and} \bibinfo{person}{Ingrid
  Verbauwhede}.} \bibinfo{year}{2021}\natexlab{}.
\newblock \showarticletitle{Systematic analysis of randomization-based
  protected cache architectures}. In \bibinfo{booktitle}{\emph{2021 IEEE
  Symposium on Security and Privacy (SP)}}. IEEE, \bibinfo{pages}{987--1002}.
\newblock


\bibitem[Qureshi(2018)]%
        {qureshi2018ceaser}
\bibfield{author}{\bibinfo{person}{Moinuddin~K Qureshi}.}
  \bibinfo{year}{2018}\natexlab{}.
\newblock \showarticletitle{CEASER: Mitigating conflict-based cache attacks via
  encrypted-address and remapping}. In \bibinfo{booktitle}{\emph{2018 51st
  Annual IEEE/ACM International Symposium on Microarchitecture (MICRO)}}. IEEE,
  \bibinfo{pages}{775--787}.
\newblock


\bibitem[Ryan(2019)]%
        {ryan2019return}
\bibfield{author}{\bibinfo{person}{Keegan Ryan}.}
  \bibinfo{year}{2019}\natexlab{}.
\newblock \showarticletitle{Return of the Hidden Number Problem.}
\newblock \bibinfo{journal}{\emph{IACR Transactions on Cryptographic Hardware
  and Embedded Systems}} (\bibinfo{year}{2019}), \bibinfo{pages}{146--168}.
\newblock


\bibitem[Sabelfeld and Myers(2003)]%
        {DBLP:journals/jsac/SabelfeldM03}
\bibfield{author}{\bibinfo{person}{Andrei Sabelfeld} {and}
  \bibinfo{person}{Andrew~C. Myers}.} \bibinfo{year}{2003}\natexlab{}.
\newblock \showarticletitle{Language-based information-flow security}.
\newblock \bibinfo{journal}{\emph{{IEEE} J. Sel. Areas Commun.}}
  \bibinfo{volume}{21}, \bibinfo{number}{1} (\bibinfo{year}{2003}),
  \bibinfo{pages}{5--19}.
\newblock
\urldef\tempurl%
\url{https://doi.org/10.1109/JSAC.2002.806121}
\showDOI{\tempurl}


\bibitem[Schindler(2015)]%
        {schindler2015exclusive}
\bibfield{author}{\bibinfo{person}{Werner Schindler}.}
  \bibinfo{year}{2015}\natexlab{}.
\newblock \showarticletitle{Exclusive exponent blinding may not suffice to
  prevent timing attacks on RSA}. In \bibinfo{booktitle}{\emph{International
  Workshop on Cryptographic Hardware and Embedded Systems}}. Springer,
  \bibinfo{pages}{229--247}.
\newblock


\bibitem[Simon et~al\mbox{.}(2018)]%
        {simon2018you}
\bibfield{author}{\bibinfo{person}{Laurent Simon}, \bibinfo{person}{David
  Chisnall}, {and} \bibinfo{person}{Ross Anderson}.}
  \bibinfo{year}{2018}\natexlab{}.
\newblock \showarticletitle{What you get is what you C: Controlling side
  effects in mainstream C compilers}. In \bibinfo{booktitle}{\emph{2018 IEEE
  European Symposium on Security and Privacy (EuroS\&P)}}. IEEE,
  \bibinfo{pages}{1--15}.
\newblock


\bibitem[Song et~al\mbox{.}(2021)]%
        {song2021randomized}
\bibfield{author}{\bibinfo{person}{Wei Song}, \bibinfo{person}{Boya Li},
  \bibinfo{person}{Zihan Xue}, \bibinfo{person}{Zhenzhen Li},
  \bibinfo{person}{Wenhao Wang}, {and} \bibinfo{person}{Peng Liu}.}
  \bibinfo{year}{2021}\natexlab{}.
\newblock \showarticletitle{Randomized last-level caches are still vulnerable
  to cache side-channel attacks! But we can fix it}. In
  \bibinfo{booktitle}{\emph{2021 IEEE Symposium on Security and Privacy (SP)}}.
  IEEE, \bibinfo{pages}{955--969}.
\newblock


\bibitem[Sung et~al\mbox{.}(2018)]%
        {sung2018canal}
\bibfield{author}{\bibinfo{person}{Chungha Sung}, \bibinfo{person}{Brandon
  Paulsen}, {and} \bibinfo{person}{Chao Wang}.}
  \bibinfo{year}{2018}\natexlab{}.
\newblock \showarticletitle{{CANAL}: a cache timing analysis framework via
  {LLVM} transformation} \emph{(\bibinfo{series}{ASE})}.
\newblock


\bibitem[Toman et~al\mbox{.}(2020)]%
        {DBLP:conf/esop/TomanSSI020}
\bibfield{author}{\bibinfo{person}{John Toman}, \bibinfo{person}{Ren Siqi},
  \bibinfo{person}{Kohei Suenaga}, \bibinfo{person}{Atsushi Igarashi}, {and}
  \bibinfo{person}{Naoki Kobayashi}.} \bibinfo{year}{2020}\natexlab{}.
\newblock \showarticletitle{ConSORT: Context- and Flow-Sensitive Ownership
  Refinement Types for Imperative Programs}. In
  \bibinfo{booktitle}{\emph{{ESOP}}} \emph{(\bibinfo{series}{Lecture Notes in
  Computer Science}, Vol.~\bibinfo{volume}{12075})}.
  \bibinfo{publisher}{Springer}, \bibinfo{pages}{684--714}.
\newblock
\urldef\tempurl%
\url{https://doi.org/10.1007/978-3-030-44914-8\_25}
\showDOI{\tempurl}


\bibitem[Vazou et~al\mbox{.}(2014)]%
        {DBLP:conf/icfp/VazouSJVJ14}
\bibfield{author}{\bibinfo{person}{Niki Vazou}, \bibinfo{person}{Eric~L.
  Seidel}, \bibinfo{person}{Ranjit Jhala}, \bibinfo{person}{Dimitrios
  Vytiniotis}, {and} \bibinfo{person}{Simon L.~Peyton Jones}.}
  \bibinfo{year}{2014}\natexlab{}.
\newblock \showarticletitle{Refinement types for Haskell}. In
  \bibinfo{booktitle}{\emph{{ICFP}}}. \bibinfo{publisher}{{ACM}},
  \bibinfo{pages}{269--282}.
\newblock
\urldef\tempurl%
\url{https://doi.org/10.1145/2628136.2628161}
\showDOI{\tempurl}


\bibitem[Vekris et~al\mbox{.}(2016)]%
        {DBLP:conf/pldi/VekrisCJ16}
\bibfield{author}{\bibinfo{person}{Panagiotis Vekris},
  \bibinfo{person}{Benjamin Cosman}, {and} \bibinfo{person}{Ranjit Jhala}.}
  \bibinfo{year}{2016}\natexlab{}.
\newblock \showarticletitle{Refinement types for TypeScript}. In
  \bibinfo{booktitle}{\emph{{PLDI}}}. \bibinfo{publisher}{{ACM}},
  \bibinfo{pages}{310--325}.
\newblock
\urldef\tempurl%
\url{https://doi.org/10.1145/2908080.2908110}
\showDOI{\tempurl}


\bibitem[Wang et~al\mbox{.}(2019a)]%
        {wang2019identifying}
\bibfield{author}{\bibinfo{person}{Shuai Wang}, \bibinfo{person}{Yuyan Bao},
  \bibinfo{person}{Xiao Liu}, \bibinfo{person}{Pei Wang},
  \bibinfo{person}{Danfeng Zhang}, {and} \bibinfo{person}{Dinghao Wu}.}
  \bibinfo{year}{2019}\natexlab{a}.
\newblock \showarticletitle{Identifying cache-based side channels through
  secret-augmented abstract interpretation}. In \bibinfo{booktitle}{\emph{28th
  $\{$USENIX$\}$ Security Symposium ($\{$USENIX$\}$ Security 19)}}.
  \bibinfo{pages}{657--674}.
\newblock


\bibitem[Wang et~al\mbox{.}(2017)]%
        {wang2017cached}
\bibfield{author}{\bibinfo{person}{Shuai Wang}, \bibinfo{person}{Pei Wang},
  \bibinfo{person}{Xiao Liu}, \bibinfo{person}{Danfeng Zhang}, {and}
  \bibinfo{person}{Dinghao Wu}.} \bibinfo{year}{2017}\natexlab{}.
\newblock \showarticletitle{Cached: Identifying cache-based timing channels in
  production software}. In \bibinfo{booktitle}{\emph{26th $\{$USENIX$\}$
  Security Symposium ($\{$USENIX$\}$ Security 17)}}. \bibinfo{pages}{235--252}.
\newblock


\bibitem[Wang et~al\mbox{.}(2019b)]%
        {wang2019time}
\bibfield{author}{\bibinfo{person}{Wubing Wang}, \bibinfo{person}{Yinqian
  Zhang}, {and} \bibinfo{person}{Zhiqiang Lin}.}
  \bibinfo{year}{2019}\natexlab{b}.
\newblock \showarticletitle{Time and Order: Towards Automatically Identifying
  $\{$Side-Channel$\}$ Vulnerabilities in Enclave Binaries}. In
  \bibinfo{booktitle}{\emph{22nd International Symposium on Research in
  Attacks, Intrusions and Defenses (RAID 2019)}}. \bibinfo{pages}{443--457}.
\newblock


\bibitem[Wang and Lee(2007)]%
        {wang2007new}
\bibfield{author}{\bibinfo{person}{Zhenghong Wang} {and}
  \bibinfo{person}{Ruby~B Lee}.} \bibinfo{year}{2007}\natexlab{}.
\newblock \showarticletitle{New cache designs for thwarting software
  cache-based side channel attacks}. In \bibinfo{booktitle}{\emph{Proceedings
  of the 34th annual international symposium on Computer architecture}}.
  \bibinfo{pages}{494--505}.
\newblock


\bibitem[Wang and Lee(2008)]%
        {wang2008novel}
\bibfield{author}{\bibinfo{person}{Zhenghong Wang} {and}
  \bibinfo{person}{Ruby~B Lee}.} \bibinfo{year}{2008}\natexlab{}.
\newblock \showarticletitle{A novel cache architecture with enhanced
  performance and security}. In \bibinfo{booktitle}{\emph{2008 41st IEEE/ACM
  International Symposium on Microarchitecture}}. IEEE,
  \bibinfo{pages}{83--93}.
\newblock


\bibitem[Weiser et~al\mbox{.}(2020)]%
        {weiser2020big}
\bibfield{author}{\bibinfo{person}{Samuel Weiser}, \bibinfo{person}{David
  Schrammel}, \bibinfo{person}{Lukas Bodner}, {and} \bibinfo{person}{Raphael
  Spreitzer}.} \bibinfo{year}{2020}\natexlab{}.
\newblock \showarticletitle{Big Numbers-Big Troubles: Systematically Analyzing
  Nonce Leakage in ($\{$EC) DSA$\}$ Implementations}. In
  \bibinfo{booktitle}{\emph{29th USENIX Security Symposium (USENIX Security
  20)}}. \bibinfo{pages}{1767--1784}.
\newblock


\bibitem[Weiser et~al\mbox{.}(2018a)]%
        {weiser2018single}
\bibfield{author}{\bibinfo{person}{Samuel Weiser}, \bibinfo{person}{Raphael
  Spreitzer}, {and} \bibinfo{person}{Lukas Bodner}.}
  \bibinfo{year}{2018}\natexlab{a}.
\newblock \showarticletitle{Single trace attack against RSA key generation in
  Intel SGX SSL}. In \bibinfo{booktitle}{\emph{Proceedings of the 2018 on Asia
  Conference on Computer and Communications Security}}.
  \bibinfo{pages}{575--586}.
\newblock


\bibitem[Weiser et~al\mbox{.}(2018b)]%
        {weiser2018data}
\bibfield{author}{\bibinfo{person}{Samuel Weiser}, \bibinfo{person}{Andreas
  Zankl}, \bibinfo{person}{Raphael Spreitzer}, \bibinfo{person}{Katja Miller},
  \bibinfo{person}{Stefan Mangard}, {and} \bibinfo{person}{Georg Sigl}.}
  \bibinfo{year}{2018}\natexlab{b}.
\newblock \showarticletitle{$\{$DATA$\}$--Differential Address Trace Analysis:
  Finding Address-based $\{$Side-Channels$\}$ in Binaries}. In
  \bibinfo{booktitle}{\emph{27th USENIX Security Symposium (USENIX Security
  18)}}. \bibinfo{pages}{603--620}.
\newblock


\bibitem[Werner et~al\mbox{.}(2019)]%
        {werner2019scattercache}
\bibfield{author}{\bibinfo{person}{Mario Werner}, \bibinfo{person}{Thomas
  Unterluggauer}, \bibinfo{person}{Lukas Giner}, \bibinfo{person}{Michael
  Schwarz}, \bibinfo{person}{Daniel Gruss}, {and} \bibinfo{person}{Stefan
  Mangard}.} \bibinfo{year}{2019}\natexlab{}.
\newblock \showarticletitle{Scattercache: Thwarting cache attacks via cache set
  randomization}. In \bibinfo{booktitle}{\emph{USENIX Security Symposium}}.
\newblock


\bibitem[Wichelmann et~al\mbox{.}(2018)]%
        {jan2018microwalk}
\bibfield{author}{\bibinfo{person}{Jan Wichelmann}, \bibinfo{person}{Ahmad
  Moghimi}, \bibinfo{person}{Thomas Eisenbarth}, {and} \bibinfo{person}{Berk
  Sunar}.} \bibinfo{year}{2018}\natexlab{}.
\newblock \showarticletitle{{MicroWalk}: A Framework for Finding Side Channels
  in Binaries}. In \bibinfo{booktitle}{\emph{ACSAC}}.
\newblock


\bibitem[Yarom and Benger(2014)]%
        {yarom2014recovering}
\bibfield{author}{\bibinfo{person}{Yuval Yarom} {and} \bibinfo{person}{Naomi
  Benger}.} \bibinfo{year}{2014}\natexlab{}.
\newblock \showarticletitle{Recovering OpenSSL ECDSA Nonces Using the FLUSH+
  RELOAD Cache Side-channel Attack.}
\newblock \bibinfo{journal}{\emph{IACR Cryptol. ePrint Arch.}}
  \bibinfo{volume}{2014} (\bibinfo{year}{2014}), \bibinfo{pages}{140}.
\newblock


\bibitem[Yarom and Falkner(2014)]%
        {yarom2014flush}
\bibfield{author}{\bibinfo{person}{Yuval Yarom} {and} \bibinfo{person}{Katrina
  Falkner}.} \bibinfo{year}{2014}\natexlab{}.
\newblock \showarticletitle{FLUSH+ RELOAD: A high resolution, low noise, L3
  cache side-channel attack}. In \bibinfo{booktitle}{\emph{23rd $\{$USENIX$\}$
  Security Symposium ($\{$USENIX$\}$ Security 14)}}. \bibinfo{pages}{719--732}.
\newblock


\bibitem[Zhang et~al\mbox{.}(2012a)]%
        {zhang2012language}
\bibfield{author}{\bibinfo{person}{Danfeng Zhang}, \bibinfo{person}{Aslan
  Askarov}, {and} \bibinfo{person}{Andrew~C Myers}.}
  \bibinfo{year}{2012}\natexlab{a}.
\newblock \showarticletitle{Language-based control and mitigation of timing
  channels}. In \bibinfo{booktitle}{\emph{Proceedings of the 33rd ACM SIGPLAN
  conference on Programming Language Design and Implementation}}.
  \bibinfo{pages}{99--110}.
\newblock


\bibitem[Zhang et~al\mbox{.}(2018)]%
        {zhang2018sc}
\bibfield{author}{\bibinfo{person}{Jun Zhang}, \bibinfo{person}{Pengfei Gao},
  \bibinfo{person}{Fu Song}, {and} \bibinfo{person}{Chao Wang}.}
  \bibinfo{year}{2018}\natexlab{}.
\newblock \showarticletitle{SC Infer: refinement-based verification of software
  countermeasures against side-channel attacks}. In
  \bibinfo{booktitle}{\emph{International Conference on Computer Aided
  Verification}}. Springer, \bibinfo{pages}{157--177}.
\newblock


\bibitem[Zhang et~al\mbox{.}(2012b)]%
        {zhang2012cross}
\bibfield{author}{\bibinfo{person}{Yinqian Zhang}, \bibinfo{person}{Ari Juels},
  \bibinfo{person}{Michael~K Reiter}, {and} \bibinfo{person}{Thomas
  Ristenpart}.} \bibinfo{year}{2012}\natexlab{b}.
\newblock \showarticletitle{Cross-VM side channels and their use to extract
  private keys}. In \bibinfo{booktitle}{\emph{Proceedings of the 2012 ACM
  conference on Computer and communications security}}.
  \bibinfo{pages}{305--316}.
\newblock


\end{thebibliography}
